\documentclass[submission, Phys]{SciPost}
\usepackage{braket}
\usepackage{tikz}
\usetikzlibrary{positioning, calc, fit, arrows.meta}
\usepackage{amsmath,amsthm,amssymb}
\usepackage{bbold}
\usepackage{bm} 
\usepackage{graphicx} 
\usepackage{comment}
\usepackage{tikz} 
\usepackage[compat=1.1.0]{tikz-feynman}

\usepackage[T1]{fontenc}
\hypersetup{
    colorlinks, 
    linkcolor={blue},
    citecolor={blue},
    urlcolor={blue}
} 

\usepackage{ulem}
\usepackage{bbm} 

\usepackage{cancel}

\begin{document}
\begin{center}{\Large \textbf{Non-Thermal Effects in Fermionic Atoms Coupled to Open Cavities}}\end{center}

\begin{center}
Jules Sueiro\textsuperscript{1}, 
M. Schir\`o\textsuperscript{1}
\end{center}

\begin{center}
{\bf 1} JEIP, UAR 3573 CNRS, Coll\`{e}ge de France, PSL Research University, 11 Place Marcelin Berthelot, 75321 Paris Cedex 05, France\\
[\baselineskip]
\end{center}



\section*{Abstract}
{\bf
We study a Fermi-Hubbard model coupled to a open dissipative single cavity mode. Using Keldysh diagrammatics we derive and solve the quantum kinetic equations for the fermions, taking the bosonic cavity mode as a source of non-equilibrium noise and dissipation. In absence of Hubbard interactions we show that the fermions reach generically a non-equilibrium steady-state, characterized by a non-thermal distribution function. Quite interestingly we demonstrate that the latter exactly nullifies the heat-current between fermions and cavity mode.
We discuss the regimes of parameters where a low-frequency effective temperature description emerges and how the fluctuations affect the mean-field phase diagram for the superradiance phase transition. Finally we include Hubbard interaction in the weak-coupling regime and show that it leads to a crossover towards a full equilibrium distribution.
}

\vspace{10pt}
\noindent\rule{\textwidth}{1pt}
\tableofcontents
\thispagestyle{fancy}
\noindent\rule{\textwidth}{1pt}
\vspace{10pt}

\section{Introduction}

The success of cavity Quantum Electrodynamics (CQED) in controlling simple quantum systems via coupling to high-finesse cavity modes~\cite{raimond2001manipulating} has spurred the interest of extending this paradigm from few body to strongly interacting collective quantum many-body systems, across different platforms in atomic physics~\cite{ritsch2012cold} and solid state~\cite{schwlawin2022cavity}.

A prominent example is provided by ultracold atoms which display a high-degree of control and tunability
and have been successfully embedded in optical resonators and coupled to optically pumped cavity modes~\cite{Mivehvar02012021}. Early demonstrations of collective phenomena in this setting include the Dicke superradiant phase transition with bosonic atoms~\cite{dimer2007proposed,baumann2010dicke,torre2013keldysh,cabllero2015quantum}. The fermionic counterpart has attracted theoretical interest~\cite{larson2008cold,piazza2014quantum,chen2014superradiance,keeling2014fermionic,piazza2014umklapp,mivehvar2017superradiant,kollath2016ultracold,Sheikhan2016cavity,ortu2026pauli,marijanovic_quench_2026} and has been recently demonstrated experimentally~\cite{zhang2021observation,helson2023density,zwettler2025cavity}. A key feature of this set up, which distinguishes it from solid-state vacuum cavity problems, is the fact that at optical frequencies the cavity mode has low losses well described by Markovian master equations. The dissipation experienced by the photon mode makes these hybrid quantum light-matter systems intrinsically open quantum many-body systems~\cite{thompsonFieldTheoryManybody2023,fazio2025manybody,sieberer2025universality}.

Recently the role of fluctuations beyond mean field on the superradiant phase transition has attracted renewed interest~\cite{bezvershenko2021dicke}. Different theoretical methods have been proposed to deal with the effect of the cavity beyond mean-field~\cite{kirton2019introduction,halati2020numerically,halati2020theoretical,jager2022lindblad,halati2025controlling,orso2025self,halati2025from}. In a recent work~\cite{tolle2025fluctuation,steadystate2026diagram} a fluctuation-induced bistability has been predicted for the Fermi-Hubbard model coupled to a cavity mode, based on the assumption that the fermionic system thermalizes effectively to a finite temperature self-consistent state via the light-matter coupling to a dissipative cavity. In this work we use Keldysh field theory and diagrammatics to study in detail the non-equilibrium steady state of fermions coupled to a cavity mode. We demonstrate that the full distribution function is highly non-thermal, yet that at low energy a well defined effective temperature can be identified, thus providing a microscopic justification to the approach of Ref.~\cite{tolle2025fluctuation}. This effective temperature can be determined by looking at the energy flow between the system and the cavity, as we explicitly demonstrate with our Keldysh approach. Furthermore we discuss how fermionic interactions contribute to the thermalization of the coupled light-matter system, leading to a crossover from a non-thermal distribution to a fully equilibrated one.

The manuscript is organized as follows: in Sec.~\ref{sec:model} we introduce the model and cast the Lindblad master equation in a Keldysh formalism. In Sec.~\ref{sec:freefermions} we describe the non-interacting limit where fermions interact and thermalize with the cavity mode. Finally in Sec.~\ref{sec:finiteU} we include the role of Fermi-Hubbard interactions on the non-equilibrium distribution of the system. In Sec.~\ref{sec:conclusions} we present our conclusions. Several Appendices complete this work.

\begin{figure}[t!]
    \centering
    \includegraphics[width=0.75\textwidth]{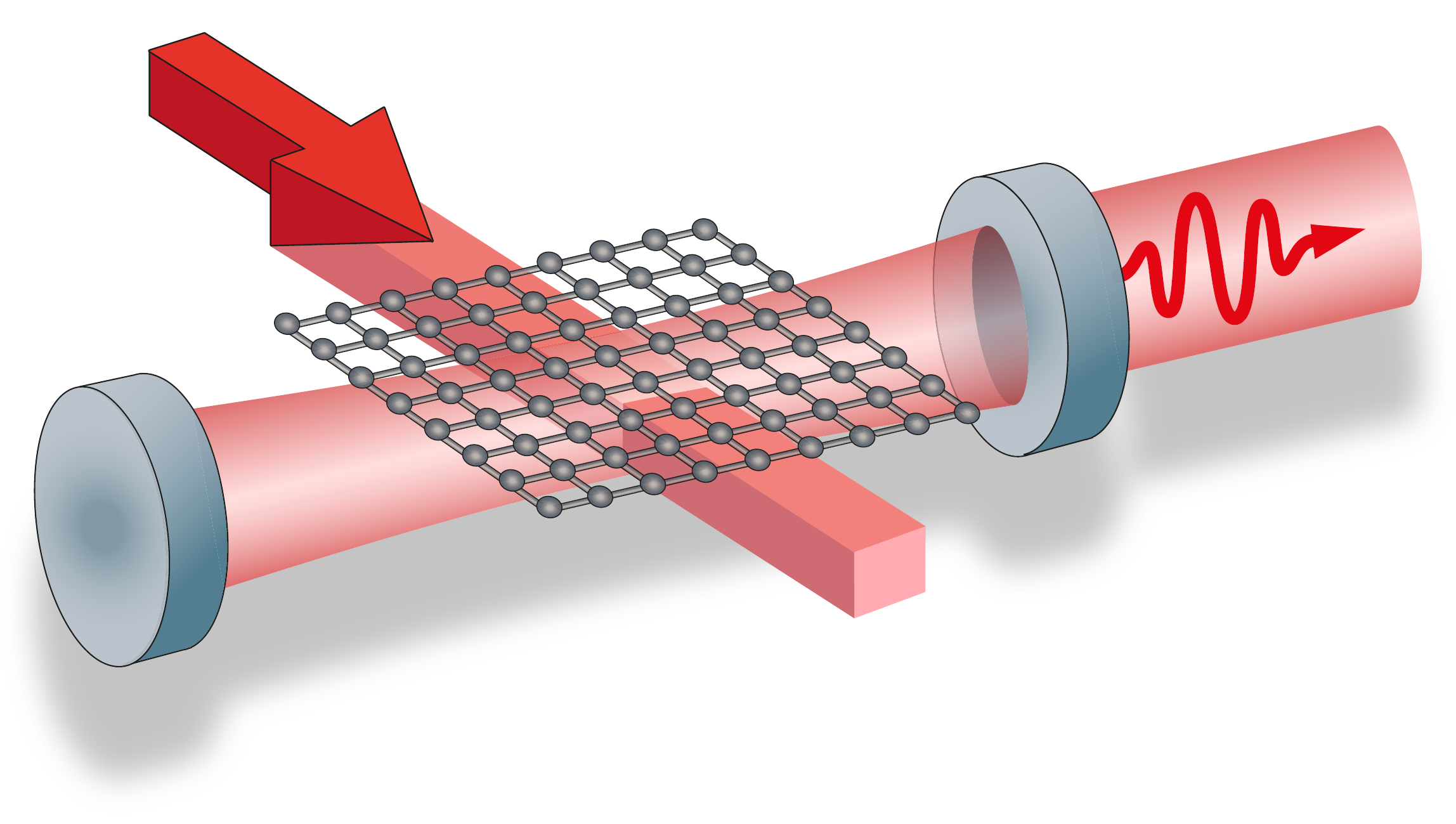}
    \caption{Illustration of the system. A system of fermionic atoms on a lattice coupled to a lossy cavity through a transverse pump. }
    \label{fig:sketch}
  \end{figure}

\section{Model and Treatment}\label{sec:model}

We consider a set of interacting fermionic atoms in an optical lattice coupled to a cavity-mode through laser assisted processes. The atoms position degree of freedom is described by the Fermi-Hubbard model via the Hamiltonian $\hat{H}_{\rm fh}$ 
\begin{align}\label{eq:HFM}
    \hat{H}_{\rm FH} &= -J \kern-0.25em\sum_{\langle i,j\rangle, \sigma} \kern-0.25em \hat c_{i,\sigma}^{\dagger}\hat c_{j,\sigma} + U \sum_{i} \hat{n}_{i, \uparrow} \hat{n}_{i, \downarrow},
\end{align}
where $J$ is the hopping rate and $U$ the local Hubbard repulsion. The cavity and the atoms are (dispersively) coupled through a transverse pump \cite{helson2023density,zwettler2025cavity} whose photons scatter into the cavity via an internal transition of the atom. In a rotating frame, the cavity is treated as a simple harmonic oscillator at the frequency $\delta = \omega_c - \omega_{\rm laser}$, i.e. $\hat{H}_{\rm ph}=\delta ~ a^{\dagger} a$, to which the fermions couple through the particle density at the wave-vector of the cavity mode~\cite{ritsch2012cold,chiriaco2022critical,tolle2025fluctuation, marijanovic_quench_2026}. We make the further simplifying assumption that the cavity mode is commensurate with the lattice spacing, so that it is appropriate to divide the system into (two) sublattices $A$ and $B$. As such, the light-matter coupling is described by the Hamiltonian 
\begin{align}\label{eq:Hlm}
\hat{H}_{\rm lm} &= \frac{g}{\sqrt{L^d}} (a+a^{\dagger})  \hat{\Delta},
\end{align}
where  $\hat{\Delta} =  \sum_{i\in A} \hat{n}_{i,  \sigma} - \sum_{i\in B}  \hat{n}_{i, \sigma} $, and the total Hamiltonian reads $\hat{H} = \hat{H}_{\rm fh} + \hat{H}_{\rm ph} + \hat{H}_{\rm lm}$. In Eq.~\eqref{eq:Hlm} we have explicitly rescaled the light-matter coupling by factor $1/\sqrt{L^d}$, which  corresponds to a light-matter coupling constant when scaling the system at a constant density of fermions (number of sites per unit volume). This choice has also the advantage of making explicitly extensive the fermion-fermion interactions obtained when the cavity is integrated out, as we will see below.

We emphasize that the basic light-matter interaction in this atomic physics realization of cavity QED differs from the one in the solid-state~\cite{dmytruk2021gauge}, where gauge invariance imposes the presence of a self-polarization term in addition to the dipole coupling in Eq.~\eqref{eq:Hlm} or a diamagnetic term.

Since we are interested in non-equilibrium effects and want to assess the thermalization of our light-matter system we include in the theoretical treatment fast Markovian losses of the cavity.  In this setting the Markovian approximation is well justified since the system is driven at finite frequency and in the rotating frame the density of states of available modes is essentially flat. The coupling to a Markovian bath, which does not satisfy fluctuation-dissipation theorem, is therefore a remaining signature of the non-equilibrium nature of the problem.
Therefore the system evolution is described by a Lindblad master equation~\cite{fazio2025manybody} for the density matrix $\rho$
\begin{align}
\partial_t \hat{\rho} = -i[\hat{H},\hat{\rho}]+\hat{L}\hat{\rho} \hat{L}^{\dagger}-\frac{1}{2}\left\{\hat{L}^{\dagger}\hat{L},\hat{\rho}\right\}
\end{align}
and the jump operator $L$ describes photon losses, $\hat{L} = \sqrt{\Gamma} ~ a$.

The model has been studied before with different methods, including mean-field methods where the light-matter interaction is decoupled or many-body adiabatic elimination methods~\cite{tolle2025fluctuation,steadystate2026diagram}. Its phase diagram
features in particular a normal phase at small $g$ and a superradiant one characterized by a coherent cavity mode and a macroscopic population of the cavity mode. 
Similar driven-dissipative superradiant phase transitions have been studied in the context of the Dicke model ~\cite{torre2013keldysh,kirton2019introduction}  where it was shown that dissipation turns the criticality from quantum to effectively classical due to the emergence of an effective temperature. However, the situation deviates from that of the Dicke model as the matter presents a broad continuum of excitation instead of the sharp resonance which only feels the cavity fluctuations at a single frequency.

\subsection{Effective Fermionic Keldysh Action}\label{sec:effective_action}

To proceed with our analysis we recast the Lindblad master equation in terms of a Keldysh action~\cite{kamenevFieldTheoryNonEquilibrium2011,siebererKeldyshFieldTheory2016,thompsonFieldTheoryManybody2023} which takes the form
\begin{equation} \label{eq:KeldyshActionInitial}
 S[\bar{\psi},\psi, \bar{\phi}, \phi] = S_{\rm fh}[\bar{\psi}, \psi] + S_{\rm ph}[\bar{\phi}, \phi] + S_{\rm lm}[\bar{\psi},\psi, \bar{\phi}, \phi] ,
\end{equation}
where $S_{\rm fh}$ is the Fermi-Hubbard action for the fermions with no dissipation, while the action for the photon is derived from the Lindblad equation as : 
\begin{equation} \label{eq:PhotonActionLindblad}
 S_{\rm ph}[\bar{\phi}, \phi] = \int dt ~ \bar{\phi}^{q} \left( \mathrm{i}  \partial_t -\delta + \mathrm{i} \Gamma/2  \right) \phi^{c} + {\rm c.c.} + \mathrm{i}{\Gamma} \bar{\phi}^{q}{\phi}^{q},
\end{equation}
So that the bare photon propagator reads :
\begin{subequations} \label{eq:PhotonPropagators}
\begin{align}
\mathcal{D}^{R/A}(\omega) &= \frac{1}{\omega - \delta \pm \mathrm{i} \Gamma/2},  \\
\mathcal{D}^K(\omega) &= \mathcal{D}^R(\omega) - \mathcal{D}^A(\omega) = \frac{-\mathrm{i} \Gamma}{\left(\omega - \delta\right)^2 + \frac{\Gamma^2}{4}} . \label{eq:PhotonPropagators_c}
\end{align}
\end{subequations}
The last term in Eq.~\eqref{eq:KeldyshActionInitial} is the light-matter contribution which is given by
\begin{align*} 
 S_{\rm lm}[\bar{\psi},\psi, \bar{\phi}, \phi] &= \int dt \left({H}_{\rm lm}^+ - {H}_{\rm lm}^-\right)=  \frac{g}{\sqrt{L^d}}  \int dt\left\{(\phi^+ \kern-0.25em + \bar{\phi}^+)  ~{\Delta}^{+} - (\phi^-\kern-0.25em + \bar{\phi}^-)  ~{\Delta}^{-}\right\}\\
 &=  \frac{g \sqrt{2}}{\sqrt{ L^d}}  \int dt \left\{ (\phi^c \kern-0.1em+ \bar{\phi}^c)  ~{\Delta}^{q} + (\phi^q\kern-0.1em + \bar{\phi}^q)  ~{\Delta}^{c}\right\},
\end{align*}
where $\Delta^{c/q} = (\Delta^{+}\pm \Delta^{-} )/2$ and we use the shorthand notation $\Delta^{\pm} \equiv \Delta[\bar{\psi}^{\pm}, \psi^{\pm}]$.

We can then integrate out exactly the cavity mode by performing the Gaussian integral in Eq.~\eqref{eq:KeldyshActionInitial} over the fields $\phi^{c,q},\bar{\phi}^{c,q}$ and obtain a fermion-only effective Keldysh action in the form
\begin{equation} \label{eq:effectiveActionWithIntegratedOutPhotons}
S_{\rm eff}[\bar{\psi},\psi] = S_{\rm fh}[\bar{\psi},\psi] - \frac{ 2 g^2}{L^d}\kern-0.5em\int_{t, t^{\prime}} \sum_{\alpha, \beta} \Delta^{\bar{\alpha}}(t)\mathcal{D}^{\alpha,\beta}(t-t^{\prime}) \Delta^{\bar{\beta}}(t^{\prime})  
\end{equation} 
The matrix notation for the photon propagator $\mathcal{D}^{\alpha,\beta}$ is related to the retarded, advanced and Keldysh components through $\mathcal{D}^{c,q} = \mathcal{D}^{R}$, $\mathcal{D}^{q,c} = \mathcal{D}^{A}$, $\mathcal{D}^{c,c} = \mathcal{D}^{K}$, while causality imposes that $\mathcal{D}^{q,q} = \mathcal{D}^{\bar{K}} = 0$ for $t\neq t^{\prime}$.

As can be seen from Eq.~\eqref{eq:effectiveActionWithIntegratedOutPhotons} the cavity mode mediates infinite range (all to all) fermion-fermion interactions, due to the single-mode nature of the cavity. Crucially, these include both coherent interactions as well as dissipative ones. 
This is most clear when the effective action is expressed back in terms of $\Delta^{\pm}$ and using of Eq.~\eqref{eq:PhotonPropagators_c} to obtain :
\begin{align} 
S_{\rm eff}[\bar{\psi},\psi] &= S_{\rm fh }[\bar{\psi},\psi] - \frac{g^2}{L^d}\int_{\omega} \left\{\Delta^+ \text{Re}~\mathcal{D}^{R}(\omega)\Delta^+ - \Delta^- \text{Re}~\mathcal{D}^{R}(\omega)\Delta^-  \right\} \label{eq:effectiveActionLindbladForm}\\ 
& + \frac{2g^2}{L^d}\int_{\omega}\mathrm{i} \left\{ \Delta^- \text{Im}~\mathcal{D}^{R}(\omega) \Delta^+ - \frac{1}{2} \Delta^+\text{Im}~\mathcal{D}^{R}(\omega)\Delta^+ - \frac{1}{2}\Delta^-\text{Im}~\mathcal{D}^{R}(\omega)\Delta^-  \right\} . \notag 
\end{align}
The first term has the Keldysh structure of an Hamiltonian contribution and describes the (retarded) coherent interaction mediated by the cavity. Meanwhile the second has the structure of a Lindblad-like dissipator and describes non-Markovian dissipation into the cavity. We note that at this stage performing a Markovian approximation on the last term would result in an effective Lindblad action with collective fermionic jump operator $\hat{L}_{\rm eff} = \frac{\sqrt{\kappa_{\rm eff}}}{\sqrt{L^d}} \hat{\Delta}$, where $ {\kappa_{\rm eff}}  = -  g^2 {\rm Im}~{\mathcal{D}}^R(0) $. One could expect in that case an infinite temperature steady-state due to the Hermitian nature of the jump. As such we see immediately that keeping the frequency dependence of the non-Markovian dissipation in Eq.~\eqref{eq:effectiveActionLindbladForm} is crucial to capture finite temperature thermalization and the correct heating exchange between atoms and cavity field.

Owing to the fact that the fermions couple through a single quadrature of the cavity, from Eq.~\eqref{eq:effectiveActionWithIntegratedOutPhotons}, the cavity-mediated interactions can be symmetrized : 
\begin{align} 
S_{\rm eff}[\bar{\psi}, \psi ]&= S_{\rm fh}[\bar{\psi}, \psi ] - \frac{2 g^2}{L^d}\int_{t, t^{\prime}} \sum_{\alpha, \beta}\frac{1}{2}\left[ \Delta^{\bar{\alpha}}(t)\mathcal{D}^{\alpha, \beta}(t-t^{\prime}) \Delta^{\bar{\beta}}(t^{\prime}) + \Delta^{\bar{\alpha}}(t^{\prime})\mathcal{D}^{\alpha, \beta}(t^{\prime}-t) \Delta^{\bar{\beta}}(t)  \right] \notag \\
&= S_{\rm fh}[\bar{\psi}, \psi ] - \frac{2 g^2}{L^d}\int_{t, t^{\prime}} \sum_{\alpha, \beta}\Delta^{\bar{\alpha}}(t) ~ \tilde{\mathcal{D}}^{\alpha, \beta}(t-t^{\prime}) ~ \Delta^{\bar{\beta}}(t^{\prime}), \label{eq:switchingToSymmetrizedInteractions}
\end{align}
where $ \tilde{\mathcal{D}}^{\alpha, \beta}(t)  \equiv  \frac{1}{2}\left[ \mathcal{D}^{\alpha, \beta}(t) +  \mathcal{D}^{\beta,\alpha}(-t) \right] $ for $\alpha, \beta = c,q$ is the symmetrized Keldysh photon propagator. In the frequency representation, this means that :
\begin{subequations} \label{eq:RAK_symmetrizedPhotonPropagator}
\begin{align}
\tilde{\mathcal{D}}^{R}(\omega) &= \frac{1}{2}\left[{\mathcal{D}}^{R}(\omega)+ {\mathcal{D}}^{A}(-\omega)\right] = \frac{\delta}{\left(\omega + \mathrm{i} \frac{\Gamma}{2}\right)^2 - \delta^2} = \left\{\tilde{\mathcal{D}}^{A}(\omega)\right\}^{\ast}, \label{eq:RAK_symmetrizedPhotonPropagator_retarded}\\
\tilde{\mathcal{D}}^{K}(\omega) &= \frac{1}{2}\left[{\mathcal{D}}^{K}(\omega)+ {\mathcal{D}}^{K}(-\omega)\right] = \frac{-\mathrm{i} \Gamma}{2}\left[\frac{1}{(\omega - \delta)^2 + \frac{\Gamma^2}{4}}+ \frac{1}{(\omega + \delta)^2 + \frac{\Gamma^2}{4}}\right] \label{eq:RAK_symmetrizedPhotonPropagator_Keldysh}.  
\end{align}
\end{subequations}
It can be noted in Eq.~\eqref{eq:RAK_symmetrizedPhotonPropagator_Keldysh} that both resonances $\omega = \pm \delta$ appear in the symmetrized Keldysh propagator.

\begin{figure}[t]
\centering
\includegraphics[width=10cm,height=6cm]{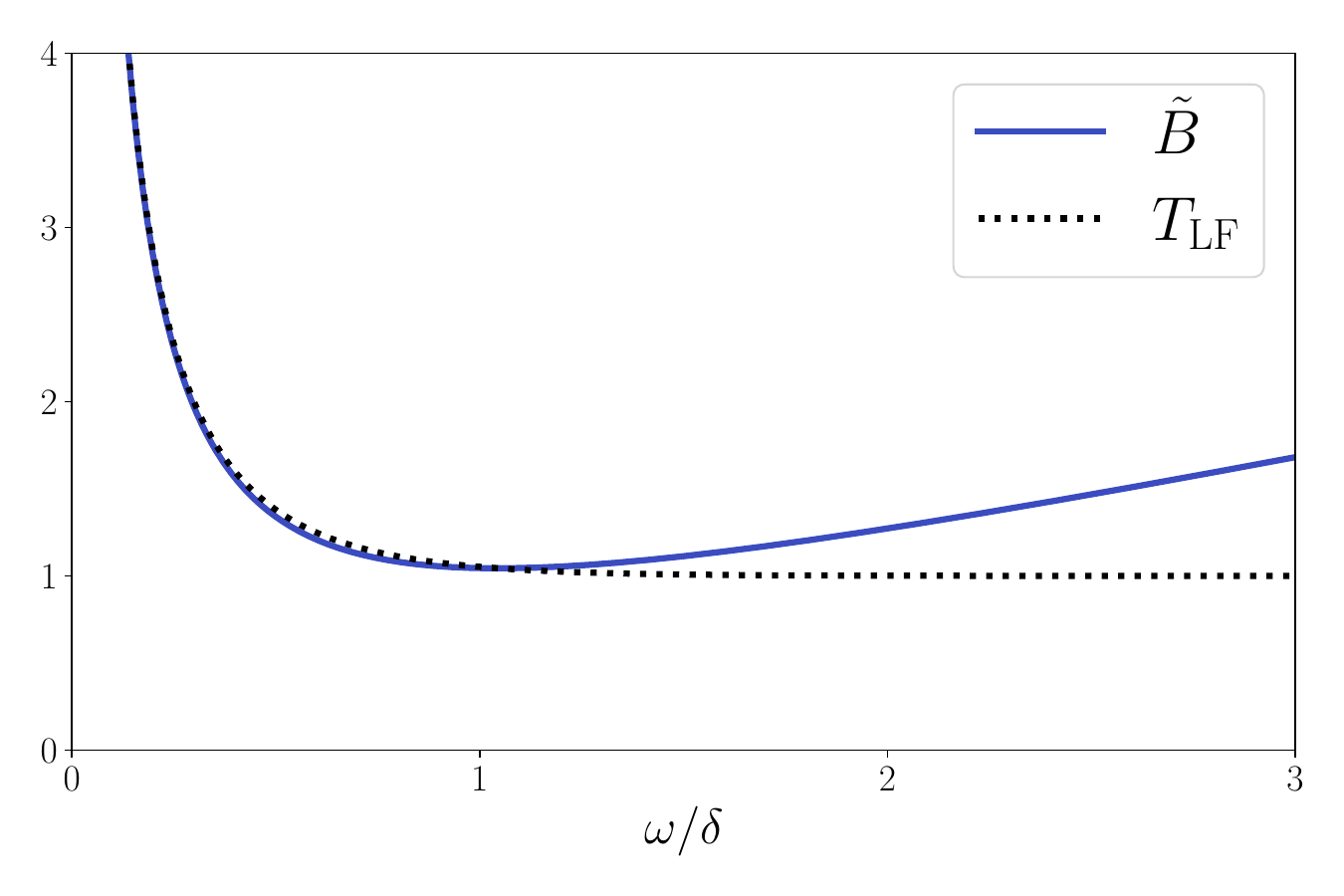}
\caption{Distribution function $\tilde{B}(\omega)$ of the symmetrized boson/cavity quadrature, compared to the thermal one
at the same low-frequency effective temperature $T_{\rm LF}$.
}
\label{fig:bosonic_distribution_function}
\end{figure}

Subsequently, one defines the distribution function $\tilde{B}(\omega)$ of the bare cavity as :
\begin{equation}\label{eq:def_DistributionFunction}
     \tilde{\mathcal{D}}^{K}(\omega) = \tilde{\mathcal{D}}^{R}(\omega) \tilde{B}(\omega) -  \tilde{B}(\omega) \tilde{\mathcal{D}}^{A}(\omega) =  \tilde{B}(\omega)\left[\tilde{\mathcal{D}}^{R}(\omega)  -  \tilde{\mathcal{D}}^{A}(\omega) \right],
\end{equation}
which in equilibrium at the inverse temperature $\beta$ would reduce to $\tilde{B}(\omega) = {\rm coth}~\beta\omega/2$ following the Fluctuation-Dissipation theorem (FDT). From the expressions of Eq.~\eqref{eq:RAK_symmetrizedPhotonPropagator}, the distribution function reads : 
\begin{equation} \label{eq:symmetrizedDistributionFunction}
\tilde{B}(\omega) = \frac{\omega^2  + \delta^2+ \frac{\Gamma^2}{4}}{2\delta \times \omega}.
\end{equation}
Furthermore, as in Ref.~\cite{torre2013keldysh, kirton2019introduction}, 
a Low Frequency Effective Temperature (LFET) $T_{\rm LF}$ arises through fitting $2/\beta\omega$ to the $\omega\longrightarrow0$ behavior of the distribution function : 
\begin{equation} \label{eq:LFET_photonQuadrature}
T_{\rm LF} =  \frac{1}{4\delta}  \left( \delta^2 + \frac{\Gamma^2}{4}\right),
\end{equation}
where have set $k_b = 1$. The distribution function $\tilde{B}$ as well as its low frequency thermal fit are displayed in Fig.~\ref{fig:bosonic_distribution_function} and clearly show that the actual distribution deviates substantially from equilibrium at frequencies $\omega \geq \delta$.

The fact that the Lindblad steady state does not describe thermal equilibrium can often be ignored since in the regime of low losses $\Gamma \ll \delta$. Indeed, the distribution function always appears together with the spectral function as in Eq.~\eqref{eq:def_DistributionFunction}, when the latter is very peaked only very few frequencies in the distribution function matter and the situation is nearly indistinguishable from thermal equilibrium at an effective temperature which may deviate from the LFET. On the contrary, in the rotating frame of the transverse laser, we consider the regime where the detuning and the losses are of the same magnitude $\delta\sim\Gamma$, and the distinction between the Lindblad steady-state and a thermal state has a significance .

After having laid out the model and formalism, it has become clear the cavity is not in a thermal state. The question of the consequences of the non-thermal cavity fluctuations on the fermions then arises.

\section{Free-Fermions Limit $U=0$}\label{sec:freefermions}

In this Section we start considering the effect of the cavity-induced coupling, setting for the moment to zero the Fermi-Hubbard interaction. In particular we compute the fermionic self-energy due to light-matter interactions and solve the kinetic equation for the fermionic distribution function. We demonstrate explicitly that the distribution function is non-thermal, yet at low energy an effective temperature can be identified that we show can be obtained by nullifying the heat flux condition.

In this limit, the spin degree of freedom is unimportant and the fermions will be treated as spinless. However, the sub-lattice index (A/B) gives rise to a $2\times2$ matrix form of the Hamiltonian, Green's function, self-energy and distribution function which will be denoted by $\check{\bullet}$ when the indices are not explicit and to which we associate the pseudo-spin Pauli matrices $\check{\tau}^{x,y,z}$. Doing so, we recover a translation invariant problem and the Green's function are expressed by their momentum $\mathbf{k}$. Finally, the Keldysh fermionic fields are rotated in the Larkin-Ovchinnikov (LO) basis $\psi^{1/2}$. As such, the Green's function are defined as
\begin{equation}\label{eq:Def_GreenFunction}
G^{ab}_{\mu\nu}(\mathbf{k},t-t')=-\mathrm{i} \langle \psi_{\mathbf{k}, \mu}^{a}(t)\bar{\psi}_{\mathbf{k}, \nu}^{b}(t')\rangle,
\end{equation}
where $a,b = 1,2$ are Keldysh indices in the LO basis and $\mu,\nu = A, B$ are the sub-lattice indices. The retarded, advanced and Keldysh components are linked to these notation through $\check{G}^R = \check{G}^{1,1}$, $\check{G}^A = \check{G}^{2,2}$, $\check{G}^K = \check{G}^{1,2}$ and $\check{G}^{2,1} = \check{G}^{\bar{K}} = 0 $ for $t\neq t^\prime$ due to causality.
We keep the notation $\hat{\bullet}$ for objects which are matrices in both the sub-lattice and Keldysh indices.  Within our pseudo spin representation of the sub-lattice indices the $\Delta$ operators are represented as : 
\begin{equation}\label{eq:expressionDeltaOperator}
    \Delta^{\alpha} = \frac{1}{2}\sum_{\mathbf{k}} \bar{\psi}^{a}_{\mathbf{k}, \mu}  \check{\tau}^{z}_{\mu,\nu} \gamma^{\bar{\alpha}}_{a,b} {\psi}^{b}_{\mathbf{k}, \nu},
\end{equation}      
where the sum over the repeated indices $\mu,\nu = A,B$  and $a,b = 1,2$ is implied, while the $2\times2$ $\gamma$ matrices \cite{kamenevFieldTheoryNonEquilibrium2011} are defined as $\gamma^c = \mathbbm{1}$ and $\gamma^{q} = \sigma^{x}$.

\subsection{Diagrammatics and Fermionic Self-Energy}

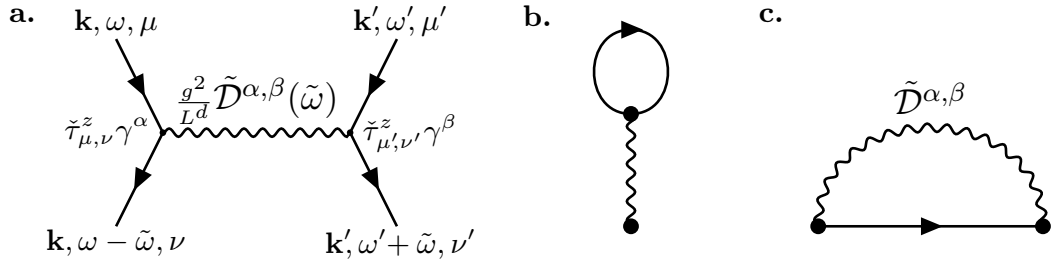
\begin{figure}[h!]
    \centering
    \resizebox{0.95\textwidth}{!}{
    \begin{tikzpicture}
        \begin{scope}
            \begin{feynman}

\vertex (inL) at (-0.5,1) ;
\vertex (inR) at (2.5, 1) ;
\vertex (coL) at (0,0) ;
\vertex (coR) at (2,0) ;
\vertex (outL) at (-0.5,-1) ;
\vertex (outR) at (2.5, -1) ;
  \diagram* {
    (inL) -- [fermion, thick]  (coL) , 
    (coL) -- [fermion, thick]  (outL) , 
    (coL) -- [boson, thick] (coR),
    (inR) -- [fermion, thick]  (coR) , 
    (coR) -- [fermion, thick]  (outR) , 
    };
\fill (coL) circle (1pt);
\fill (coR) circle (1pt);
\node[anchor = center] at (1, 0.35) { ${\scriptstyle\frac{g^2}{L^d}}\tilde{\mathcal{D}}^{\alpha, \beta}(\tilde{\omega})$}; 
\node[anchor=east] at (0, 0) { \footnotesize $\check{\tau}^{z}_{\mu,\nu} \gamma^{\alpha}$ }; 
\node[anchor=west] at (2, 0) { \footnotesize $\check{\tau}^{z}_{\mu^{\prime}\kern-0.25em,\nu^{\prime}} \gamma^{\beta}$ }; 
\node[anchor=south] at (-0.5, 0.9) { \footnotesize $\mathbf{k}, \omega, \mu$ }; 
\node[anchor=north] at (-0.5, -0.9) { \footnotesize $\mathbf{k}, \omega - \tilde{\omega}, \nu$ }; 
\node[anchor=south] at (2.5, 0.9) { \footnotesize $\mathbf{k}^{\prime}\kern-0.25em, \omega^{\prime}\kern-0.25em, \mu^{\prime}\kern-0.25em$ }; 
\node[anchor=north] at (2.5, -0.9) { \footnotesize $\mathbf{k}^{\prime}\kern-0.25em, \omega^{\prime}\kern-0.25em + \tilde{\omega}, \nu^{\prime}\kern-0.25em$ }; 
\end{feynman}
\end{scope}

\begin{scope}[xshift = 5cm, yshift = -1 cm, scale = 1.2]
\begin{feynman}
\vertex (a) at (0,0) ;
\vertex (b) at (0,1)  ;
\vertex (c) at (0,1.75)  ;
  \diagram* {
    (a) -- [boson,   thick]  (b) , 
    (b) -- [   thick, half left, with arrow=0.999]  (c) , 
    (c) -- [   thick, half left]  (b) , 
    };
\fill (a) circle (2pt);
\fill (b) circle (2pt);
\end{feynman}
\end{scope}

\begin{scope}[xshift = 7cm, yshift = -1 cm, scale = 1.2]
\begin{feynman}
\vertex (a) at (0,0) ; 
\vertex (b) at (2,0)  ; 
  \diagram* {
    (a) -- [boson,   thick,  edge label = \(\tilde{\mathcal{D}}^{\alpha,\beta}\), half left]  (b) , 
    (a) -- [fermion,   thick]  (b) , 
    };
\fill (a) circle (2pt);
\fill (b) circle (2pt);
\end{feynman}
\end{scope}

\node at (-1.5,1.25) {\footnotesize\textbf{a.}};
\node at (4,1.25) {\footnotesize\textbf{b.}};
\node at (6.5,1.25) {\footnotesize\textbf{c.}};

    \end{tikzpicture}
    }
 
    \caption{\textbf{a.} Convention and notations for the vertex cavity mediated interactions. \textbf{b.} Hartree contribution to the cavity-mediated self-energy which capture the effect of mean-field between the cavity and matter. \textbf{c.} Fock contribution to the cavity-mediated self-energy representing the effect of the (non-thermal) cavity fluctuations on the fermions. To preserve energy conservation in this non-equilibrium setting, all the diagrammatic expansion are bold i.e. the fermion lines appearing in diagrams are dressed propagators.}
    \label{fig:CavityMediatedInteractions}
\end{figure}

Starting from the symmetrized interactions from Eq.~\eqref{eq:switchingToSymmetrizedInteractions}, the cavity mediated vertex is be represented diagrammatically in Fig.~\ref{fig:CavityMediatedInteractions}.

Notably, the all-to-all interactions do not transfer any momentum and have a $g^2/L^d$ prefactor. Consequently, only diagrams with as many free-$k$ summations as there are photon lines contribute in the thermodynamic limit, which imposes clearly dominant diagrams and sub-leading contributions. The single-particle properties of the system coupled to the cavity are all contained in the Green's function. The effect of the cavity on the fermions is then captured through the Dyson equation which characterizes how the fermions' propagator is dressed from the action of the cavity-mediated interactions. In the steady state where time translation invariance is recovered, it takes the well-known form : 
\begin{equation}\label{eq:Def-dysonEquation}
    \hat{G}^{-1}(\mathbf{k}, \omega) = \hat{G}_0^{-1}(\mathbf{k}, \omega) -  \hat{\Sigma}(\mathbf{k}, \omega),
\end{equation} 
where the $\bullet^{-1}$ has to be understood as a matrix inversion for both the sublattice and Keldysh indices, while $\hat{G}$ is the dressed Green's function defined in Eq.~\eqref{eq:Def_GreenFunction} and $\hat{G}_0$ is the bare Green's function. Finally, the self-energy $\hat{\Sigma}$ is the sum of all one particle irreducible connected diagrams i.e. which cannot be split into two parts when cutting a single fermion line.

The leading contribution to the self-energy is the Hartree diagram shown in Fig.~\ref{fig:CavityMediatedInteractions}b. and which corresponds to the mean-field contribution. It can be expressed following the Feynman rules of Fig.~\ref{fig:CavityMediatedInteractions}a. as :
\begin{align*}
    \check{\Sigma}^{a,b}_{\rm H} = \mathrm{i} \check{\tau}^z \gamma^{\alpha}_{a,b}\mathcal{D}^{\alpha,\beta}(\tilde{\omega} = 0) \gamma^{\beta}_{c,d} \frac{g^2}{L^d}\sum_{\mathbf{k}} \int {\rm tr} ~\check{\tau}^z \check{G}^{c,d}(\mathbf{k},\omega) ~ d\omega,
\end{align*}
and is frequency independent i.e. time-local.

The expression is further simplified using the constraints of causality and commutation relations on Green's function, leaving only the retarded photon propagator and the Keldysh component of the fermions Green's function. Identifying : 
\begin{equation}\label{eq:expectationValueDeltaFromGK}
    \langle \Delta\rangle = \frac{\mathrm{i}}{2}\sum_{\mathbf{k}} \int {\rm tr}~\check{\tau}^z \check{G}^{K}(\mathbf{k},\omega) ~ d\omega,
\end{equation}
the Hartree self-energy appears as : 
\begin{equation}
    \hat{\Sigma}_{\rm H}= - g^2 \frac{ 2\delta}{\delta^2 + \frac{\Gamma^2}{4} } \frac{\left<\Delta\right>}{L^d}  \check{\tau}^z \gamma^c .
\end{equation}

This contribution to the self-energy can be absorbed into the Hamiltonian via : 
\begin{equation}\label{eq:meanFieldHamiltonian}
    \hat{H}_{\rm fh} \longmapsto \hat{H}_{\rm fh + H} =  \hat{H}_{\rm fh} - g^2 \frac{ 2 \delta}{\delta^2 + \frac{\Gamma^2}{4} } \frac{\left<\Delta\right>}{L^d} \hat{\Delta},
\end{equation}
so that the mean-field equations arise from self-consistently satisfying Eq.~\eqref{eq:expectationValueDeltaFromGK} with the Green's function resulting from the shifted Hamiltonian. The mean-field equations capture the phase transition at a critical light-matter coupling $g_{\rm c }$ between the normal phase and the superradiant phase. The latter is characterized by a macroscopic occupation of the cavity $\phi = \langle a^\dagger + a\rangle/ \sqrt{L^d} \neq 0 $ together with a staggered density of fermions in space $\langle\Delta\rangle\neq 0$.

The mean-field result is usually understood as a decoupling of the light and matter sectors. Then, the light sees a coherent source $\hat{H}_{\rm MF, L} = {g} \langle\Delta\rangle ( a^{\dagger} + a)/\sqrt{L^d}$, while the fermions experience a potential given by $\hat{H}_{\rm MF, M}= {g} \langle a^{\dagger} + a\rangle \hat{\Delta}/ \sqrt{L^d}$. This picture is equivalent to Eq.~\eqref{eq:meanFieldHamiltonian}. Indeed, Gaussian integration shows that under the coherent source the cavity quadrature develops an expectation value $ \langle a^{\dagger} + a\rangle = 2\tilde{\mathcal{D}}^{R}(\tilde{\omega} = 0){g}\langle\Delta\rangle/ \sqrt{L^d}$, which, when substituted in $\hat{H}_{\rm MF, M}$, reproduces the effective matter Hamiltonian in Eq.~\eqref{eq:meanFieldHamiltonian}.

Most notably, the leading contribution to the self energy leaves the Keldysh sector -and hence the occupations- unchanged. Going to the first sub-leading contribution shown in Fig.~\ref{fig:CavityMediatedInteractions}c. solves this issue. It corresponds to the Fock self-energy and has the expression :
\begin{equation}\label{eq:FockSelfEnergy_KeldyshIndices}
    \check{\Sigma}_{\rm F}^{a,d}(\mathbf{k}, \omega) = \mathrm{i} \frac{g^2}{L^d} \int d\tilde{\omega} ~ \tilde{\mathcal{D}}^{\alpha, \beta}(\tilde{\omega}) \gamma^{\alpha}_{a,b} ~\check{\tau}^{z} ~\check{G}^{b,c}(\mathbf{k}, \omega - \tilde{\omega}) ~\gamma^{\beta}_{c,d}  ~ \check{\tau}^{z}.
\end{equation}
The Keldysh index can be contracted and yield the expressions : 
\begin{subequations} \label{eq:self-energyGeneralExpression_detail_tau}
\begin{align}
\check{\Sigma}^{R/A}_{\rm F}(\mathbf{k}, \omega) &= \mathrm{i} \frac{g^2}{L^d} \kern-0.25em \int_{\tilde{\omega}} \kern-0.25em \left\{ \tilde{\mathcal{D}}^{K}(\tilde{\omega}) \check{\tau}^{z}\check{G}^{R/A}(\mathbf{k}, \omega - \tilde{\omega})  \check{\tau}^{z}\kern-0.25em  + \tilde{\mathcal{D}}^{R/A}(\tilde{\omega}) \check{\tau}^{z} \check{G}^{K}(\mathbf{k}, \omega - \tilde{\omega})   \check{\tau}^{z}  \right\} ,  \label{eq:self-energyGeneralExpression_detailRA_tau}\\ 
\check{\Sigma}^{K}_{\rm F}(\mathbf{k}, \omega)& = \mathrm{i} \frac{g^2}{L^d} \kern-0.25em \int_{\tilde{\omega}} \kern-0.25em  \left\{ \tilde{\mathcal{D}}^{K}(\tilde{\omega}) \check{\tau}^{z} \check{G}^{K}(\mathbf{k}, \omega - \tilde{\omega})   \check{\tau}^{z} \kern-0.25em +  [\tilde{\mathcal{D}}^{R}(\tilde{\omega}) - \tilde{\mathcal{D}}^{A}(\tilde{\omega})]  \check{\tau}^{z} [\check{G}^{R}(\mathbf{k}, \omega - \tilde{\omega}) \right.\notag \\ &\left.  - \check{G}^{A}(\mathbf{k}, \omega - \tilde{\omega}) ] \check{\tau}^{z} \right\} ,  \label{eq:self-energyGeneralExpression_detailK_tau}
\end{align}
\end{subequations}
where causality relations $\tilde{\mathcal{D}}^R(t) G^A(t) = \tilde{\mathcal{D}}^A(t) G^R(t) = 0 $  have been used in the second equation to express the last terms with the spectral functions of the photon and fermions.

Having incorporated the mean-field into the single-particle Hamiltonian, the Fock self-energy captures the effect of the cavity fluctuations around the mean-field. The associated Dyson equation is solved by mapping it into a kinetic equation, as detailed below.

\subsection{Kinetic Equation}

In equilibrium, the knowledge of the spectral properties, stored in the retarded sector, is enough to fully characterize the occupations via the fluctuation-dissipation theorem. However, out of equilibrium, one needs to solve for both the spectral function and the occupations of the state. The effect of the non-thermal cavity fluctuations on the fermions is book-kept by the Keldysh sector. To separate the occupations from the spectrum, similarly to what has been done for the boson in Eq.~\eqref{eq:def_DistributionFunction}, the fermions distribution function $\check{F}(\mathbf{k}, \omega )$ is defined through : 
\begin{equation}\label{eq:definitionFermionDistributionFunctions} 
    \check{G}^{K}(\mathbf{k}, \omega ) = \check{G}^R(\mathbf{k}, \omega )\check{F}(\mathbf{k}, \omega ) -  \check{F}(\mathbf{k}, \omega )\check{G}^A(\mathbf{k}, \omega ).
\end{equation}
Starting from the Dyson equation from Eq.~\eqref{eq:Def-dysonEquation}, we will derive an equation on the distribution function and through which we solve for the non-thermal steady state of the system.

The Keldysh structure of the Dyson equation \eqref{eq:Def-dysonEquation} can be unfolded into equations for the retarded/advanced and the Keldysh components of the Green's function : 
\begin{subequations} \label{eq:dysonEquationCheckMatrices}
\begin{align}
\left(\left[\check{G}^R_0\right]^{-1} - \check{\Sigma}^R\right) \check{G}^R &= 1,  \label{eq:dysonEquationCheckMatrices_R}\\
\left(\left[\check{G}^R_0\right]^{-1} - \check{\Sigma}^R\right) \check{G}^K - \check{\Sigma}^K \check{G}^A &= 0  \label{eq:dysonEquationCheckMatrices_K},
\end{align}
\end{subequations}
where both the $\mathbf{k}$ and $\omega$ dependence are implied. The single-particle Hamiltonian is defined via : 
\begin{equation} \label{eq:defh_k}
\check{h}_{\mathbf{k}} = J ( 1 + \cos k_{\parallel} ) ~ \check{\tau}^x + J \sin k_{\parallel} ~\check{\tau}^y + g  \phi ~ \check{\tau}^z  +  \left(\varepsilon_{\perp}(\mathbf{k}_{\perp}) - \mu\right)\mathbbm{1},
\end{equation}
where $\phi = 2 g\tilde{\mathcal{D}}^{R}(\omega = 0) \langle \Delta \rangle/ L^d = \langle a^{\dagger} + a\rangle/\sqrt{L^d}$ and $\varepsilon_{\perp}(\mathbf{k}_{\perp})$ is the single-particle dispersion in the remaining dimensions of the lattice, perpendicular to the direction of the staggered potential. It has to be noted that the Hartree (mean-field) contribution has been included in the single-particle Hamiltonian. Furthermore, this matrix can be diagonalized to find the (mean-field) energies : 
\begin{equation}\label{eq:MF_singlePartcileEnergies}
    \varepsilon_{\mathbf{k}, \nu} = \nu \sqrt{2 J^2 \left(1+\cos(k_{\parallel}) \right)+ g^2 \phi^2} + \varepsilon_{\perp}(\mathbf{k}_{\perp}) - \mu, 
\end{equation}
for $\nu = \pm$. This way the bare Green's function takes the form : 
\begin{equation} \label{eq:tau_mapping_GF}
\check{G}^{R/A}_0(\mathbf{k}, \omega ) = \left(\omega - \check{h}_{\mathbf{k}} \pm \mathrm{i} 0^+ \right)^{-1}.
\end{equation}
 Then, the distribution function parametrization of the Keldysh Green's function~\eqref{eq:def_DistributionFunction} as well as the Dyson equation in the Keldysh sector~\eqref{eq:dysonEquationCheckMatrices_K} yields the kinetic equation for the steady state distribution function :  
\begin{equation}\label{eq:KineticEquationSteadyState}
-\left[ \check{h}_{\mathbf{k}} ~ ,~ \check{F}(\mathbf{k}, \omega ) \right] = \check{\Sigma}^K(\mathbf{k}, \omega ) - \left(\check{\Sigma}^R(\mathbf{k}, \omega ) ~ \check{F}(\mathbf{k}, \omega ) - \check{F}(\mathbf{k}, \omega )~ \check{\Sigma}^A(\mathbf{k}, \omega )\right)
\equiv  \check{\mathcal{I}}_{\mathbf{k}}^{\rm cav}[\check{F} ], 
\end{equation}
where we have introduced the collision integral $\check{\mathcal{I}}_{\mathbf{k}}^{\rm cav}[\check{F} ]$.

At this point, the retarded self-energy  $\check{\Sigma}^R$ is separated into its Hermitian ( $\sim$ real) and anti-Hermitian ( $\sim$ imaginary) parts  : 
\begin{subequations} \label{eq:effectiveHamiltonian_and_dissipation}
\begin{align}
\delta\check{h}_{\mathbf{k}}(\omega) &\equiv \check{\Sigma}^R (\mathbf{k}, \omega) + \check{\Sigma}^A(\mathbf{k}, \omega) ,\\
\delta\check{Q}_{\mathbf{k}}(\omega) &\equiv \check{\Sigma}^R (\mathbf{k}, \omega) - \check{\Sigma}^A(\mathbf{k}, \omega).
\end{align}
\end{subequations}
With those notations, the kinetic equation is rewritten as : 
\begin{equation} \label{eq:kineticEquation_freeFermions_V2}
\left[  \check{h}_{\mathbf{k}}  + \delta\check{h}_{\mathbf{k}}(\omega) ~ , ~ \check{F}(\mathbf{k}, \omega) \right]  = \check{\Sigma}^K(\mathbf{k}, \omega) - \frac{1}{2} \left\{ \delta\check{Q}_{\mathbf{k}}(\omega) ~,~ \check{F}(\mathbf{k}, \omega)  \right\},
\end{equation}
where $\{ \bullet , \bullet\}$ denotes the anti-commutator.

Recall that $\Sigma^{R/A/K} \sim g^2 / L^d$, and as such is sub-dominant in the thermodynamic limit. As such, $\delta\check{h}_{\mathbf{k}}(\omega)$ is considered to be negligible with respect to $\check{h}_{\mathbf{k}}$ and thus the shift in the single-particle spectrum due to the cavity mediated interaction is neglected. However, the $\delta\check{Q}_{\mathbf{k}}(\omega)$ in the r.h.s. of Eq.~\eqref{eq:kineticEquation_freeFermions_V2}, which quantifies the lifetimes due to cavity mediated interactions, is kept as the only thing to compare them to is $\check{\Sigma}^K(\mathbf{k}, \omega)$ which is also of order $g^2 / L^d$.
Furthermore, from Eq.~\eqref{eq:kineticEquation_freeFermions_V2} it can be argued that the coherence between the two bands are negligible. Indeed, the right hand side of Eq.~\eqref{eq:kineticEquation_freeFermions_V2} is very small and thus the distribution function can be satisfactorily considered to commute with the single-particle Hamiltonian. As such, the right hand side can be projected onto the eigen-spaces of the single-particle Hamiltonian since any deviation can just be absorbed by a counter term of order $\sim g^2/L^d$ in the distribution function. Finally, the imaginary part of the self-energy is small and the quasi-particle peaks in the spectral function can be considered as Dirac deltas. Under those assumptions, the Keldysh Green's function is expected to be well captured by the ansatz: 
\begin{equation} \label{eq:onShellCommutingAnsatz}
\check{G}^{K}(\mathbf{k}, \omega) = - \mathrm{i} 2\pi \sum_{\nu = \pm} f_{\mathbf{k}, \nu}~ \delta(\omega - \varepsilon_{\mathbf{k}, \nu} )~ \check{P}_{\mathbf{k}, \nu},
\end{equation}
where $\check{P}_{\mathbf{k}, \nu}$ are the projectors on the eigenvectors $\check{h}_{\mathbf{k} }$ with the associated eigenvalue $\varepsilon_{\mathbf{k}, \nu} $ for $\nu = \pm$. While $f_{\mathbf{k}, \nu}$ is related to the distribution function $\check{F}$ defined in Eq.~\eqref{eq:definitionFermionDistributionFunctions} via $\check{F}(\mathbf{k}, \omega) = \sum_{\nu} f_{\mathbf{k}, \nu}  \check{P}_{\mathbf{k}, \nu}$.

As detailed in Appendix~\ref{app:DerivationOfKineticEquation}, making use of the ansatz above leads to a collision integral of the form : 
\begin{equation}\label{eq:collisionIntegralForm}
   \kern-0.175em \mathcal{I}_{\mathbf{k}, \nu}^{\rm cav}\kern-0.1em = \kern-0.1em\frac{g^2}{2L^d} \kern-0.1em\left(1 \kern-0.1em- \kern-0.1em u_{\mathbf{k}, z}^2\right) \kern-0.1em\nu {\rm Im}\tilde{\mathcal{D}}^{R}(\Delta \varepsilon_{\mathbf{k}}) \left\{ \nu\tilde{B}( \Delta \varepsilon_{\mathbf{k}})    f_{\bar{\nu}} (\mathbf{k}) \kern-0.1em +\kern-0.1em 1  \kern-0.1em-\kern-0.1em \left( \nu  \tilde{B}(\Delta \varepsilon_{\mathbf{k}})  \kern-0.1em + \kern-0.1em f_{\bar{\nu}}(\mathbf{k}) \kern-0.1em  \right)\kern-0.1em  f_{\nu}(\mathbf{k})\right\}\kern-0.1em, 
\end{equation}
where $u_{\mathbf{k}, z} =  g\phi/\sqrt{2J^2 (1+ \cos k_{\parallel}) + g^2 \phi^2} $ is the $z$-component of the (normalized) Hamiltonian in the Pauli matrices basis, see Eq.~\eqref{eq:defh_k}. In the expression above we note that the cavity distribution and spectral functions, $\tilde{B}(\omega)$ and ${\rm Im}\tilde{\mathcal{D}}^{R}(\omega)$ defined in  Sec.~\ref{sec:effective_action}, are evaluated at $\omega=\Delta\varepsilon_\mathbf{k} \equiv \varepsilon_{\mathbf{k},+} -\varepsilon_{\mathbf{k},-}$.

The steady-state condition is obtained by setting to zero the 
collision integral, thus imposing :
\begin{equation} \label{eq:kineticEquationOnShellFinal}
1 -\nu \tilde{B}(\Delta \varepsilon_{\mathbf{k}}) \left[f_{\mathbf{k},\nu}  - f_{\mathbf{k}, \bar{\nu}} \right]   - f_{\mathbf{k},\nu} f_{\mathbf{k},\bar{\nu}} =0.
\end{equation}
However, this equation only fixes $f_{\mathbf{k}, + } - f_{\mathbf{k}, - } $ due to the fact that the cavity doesn't scatter between different $\mathbf{k}$ and cannot change the total population of each $\mathbf{k}$ controlled by $f_{\mathbf{k}, + } + f_{\mathbf{k}, - } = 2( 1 - \bar{n}) $ where $\bar{n} = \langle n_A\rangle + \langle n_B\rangle$ is the number of electrons per unit cell.

In the case of half-filling $\bar{n} = 1$, and in particular in 1D, we simply get $f_{\mathbf{k}, + } = - f_{\mathbf{k}, - }$, so that the equation simplifies into :
\begin{equation*}
    1 - 2\tilde{B}(\Delta \varepsilon) f_{\mathbf{k}, +} - f_{\mathbf{k}, +}^2 =0,
\end{equation*} 
for which the solution is found to be :
\begin{equation} \label{eq:finalFormOnshellDistribution}
f_{\mathbf{k}, +} = \tilde{B}(\Delta\varepsilon_{\mathbf{k}}) - \sqrt{\tilde{B}(\Delta\varepsilon_{\mathbf{k}})^2 - 1}.
\end{equation} 
Moreover, particle-hole symmetry further implies that $ \Delta\varepsilon_{\mathbf{k}} = 2\varepsilon_{\mathbf{k}, +}$ which allows to write the occupation of a mode as functions of its energy :
\begin{equation}\label{eq:occupationAsEnergy}
    f_{\mathbf{k}, \nu} = \tilde{F}(\varepsilon_{\mathbf{k}, \nu}),
\end{equation}
where the function $\tilde{F}(\varepsilon)$ is given by :  
\begin{equation*}
    \tilde{F}(\varepsilon) = \tilde{B}(2\varepsilon) - {\rm sgn}(\varepsilon)\sqrt{\tilde{B}(2\varepsilon)^2 - 1},
\end{equation*}
We note that at this level of approximation the fermionic distribution function is completely controlled by the bosonic one, $\tilde{B}(2\varepsilon)$ defined in Sec.~\ref{sec:effective_action} and does not depend neither on other fermionic parameters nor on the light matter coupling $g$. The latter only appears as a prefactor of the collision integral thus setting the relaxation time towards the steady state.

At this point, using the formula $\tanh(a) = {\rm coth}(2a)- \sqrt{{\rm coth}(2a)^2 - 1}$ (for $a>0$), we verify that if the bosons are thermal at a temperature $T_0$ then so are the fermions at the same temperature $T_0$.

\subsection{Results: fermionic distribution and effective temperature}

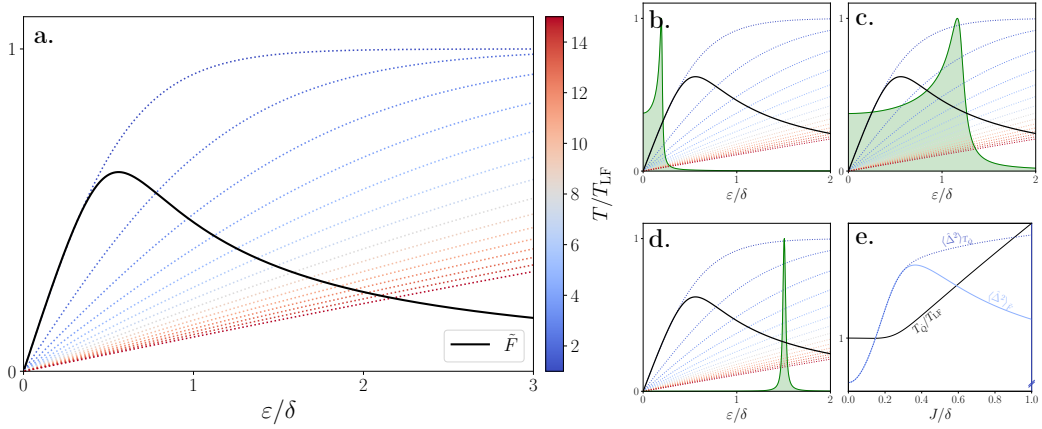
\begin{figure}[t]
\centering
\resizebox{0.95\textwidth}{!}{
\begin{tikzpicture}
\begin{scope}[xshift = 0cm, yshift = 0cm, scale = 1]
\node at (0,0){\includegraphics{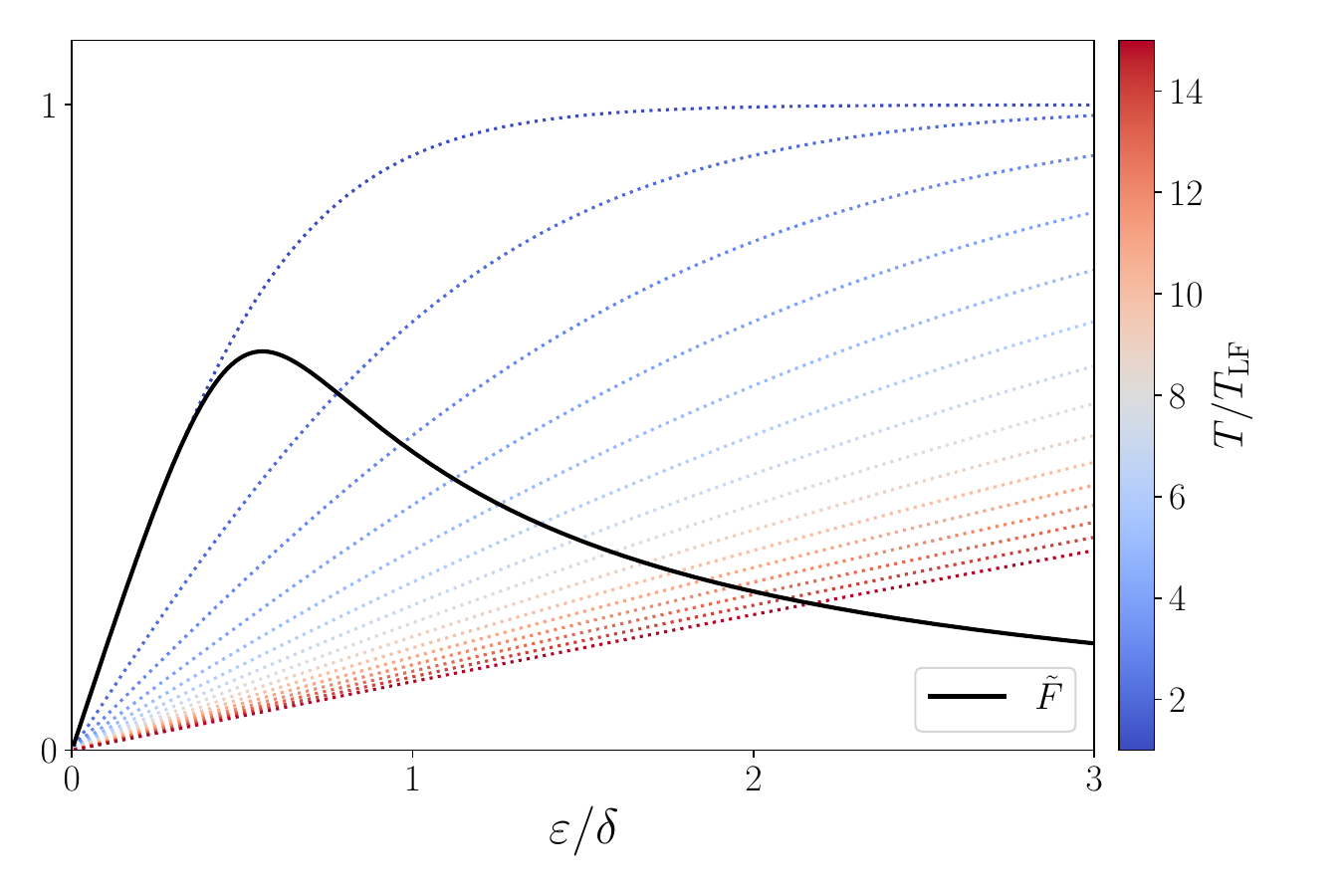}  };
\end{scope}
\begin{scope}[xshift = 14cm, yshift = 4cm]
\node at (0,0){\includegraphics[scale = 0.6]{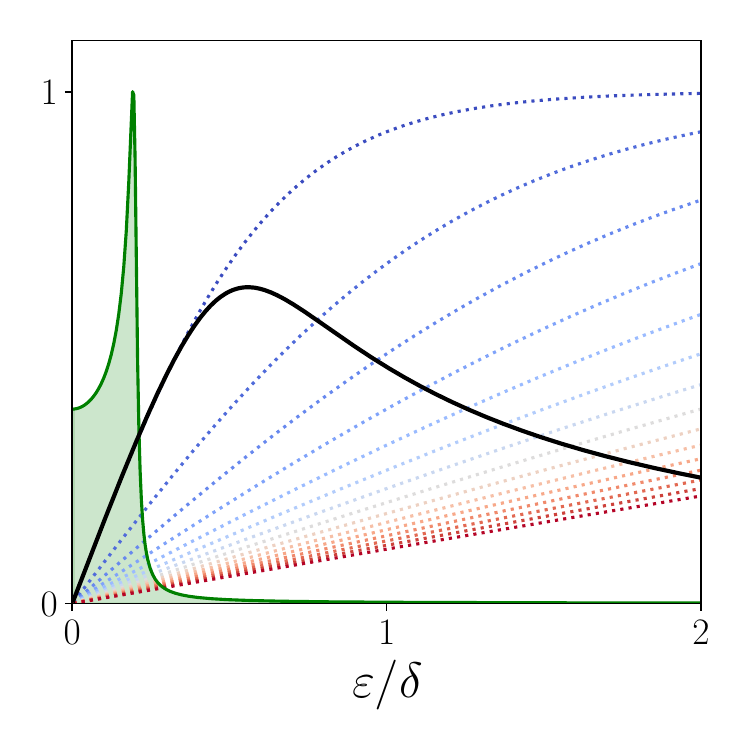}  };
\end{scope}
\begin{scope}[xshift = 14cm, yshift = -3.5cm]
\node at (0,0){\includegraphics[scale = 0.6]{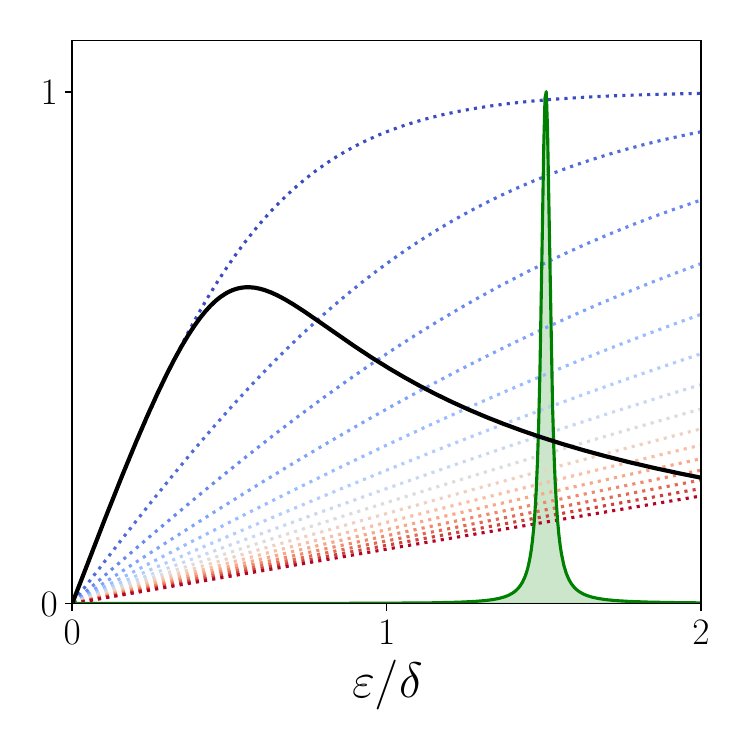}  };
\end{scope}
\begin{scope}[xshift = 21cm, yshift = 4cm]
\node at (0,0){\includegraphics[scale = 0.6]{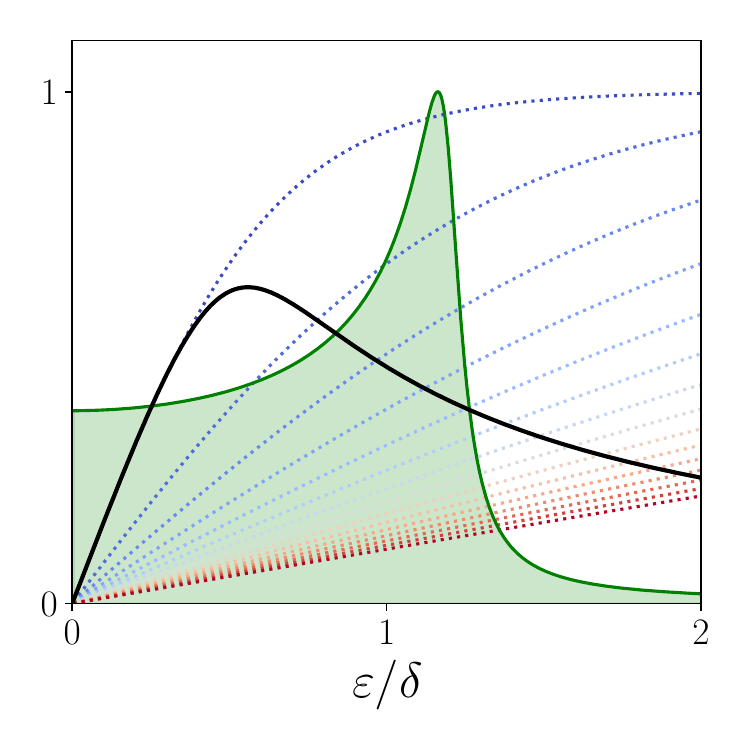}  };
\end{scope}
\begin{scope}[xshift = 21cm, yshift = -3.5cm]
\node at (0,0){ \includegraphics[scale = 0.6]{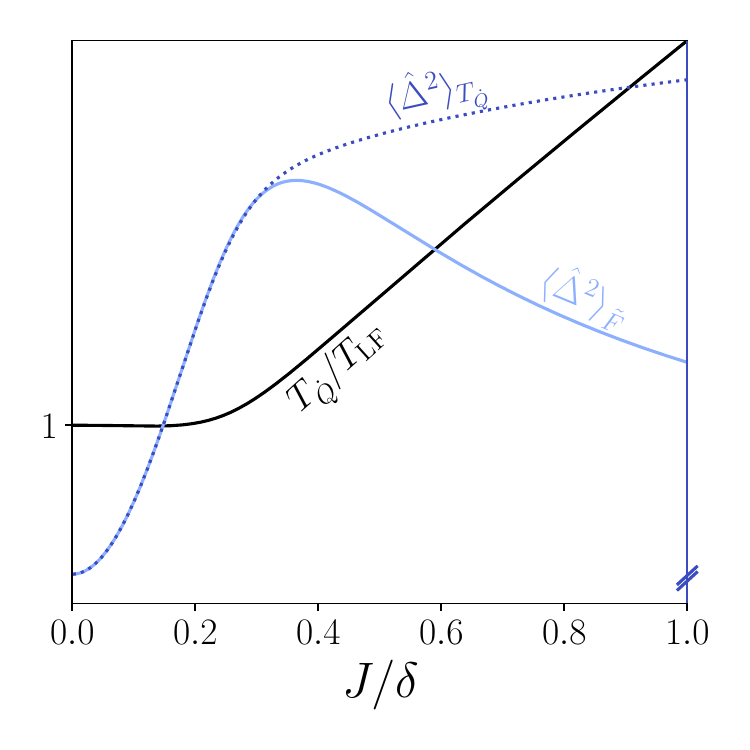} };
\end{scope}
\node at (-9.5, 6.25) {\Huge\textbf{a.}};
\node at (11.5, 6.85) {\Huge\textbf{b.}};
\node at (18.5, 6.85) {\Huge\textbf{c.}};
\node at (11.5, -.6) {\Huge\textbf{d.}};
\node at (18.5, -.6) {\Huge\textbf{e.}};
\end{tikzpicture}
}
\caption{\textbf{a.} Non-equilibrium distribution function of the fermions $\tilde{F}(\varepsilon)$ as a function of the quasi-particle energy in units of the cavity frequency $\delta$ (full line). The dotted lines show the would-be equilibrium distribution function of fermions for different temperature in units of the LFET $T_{\rm LF}$.   \textbf{b.} Distribution function superimposed with the density of states (green) in the normal phase for $\delta = 10 J$. \textbf{c.} for $\delta = 2 J$. \textbf{d.} in the superradiant phase with $g\phi = 1.5 \delta$ and $\delta = 5J$. \textbf{e.} In black, the effective temperature $T_{\dot{Q}}$ -in units of $T_{\rm LF}$- such that the heat-flux from the fermions to the cavity vanishes ${\dot{Q}} = 0 $, as a function of the bandwidth of the fermions $J/\delta$. In blue, the fluctuations of $\hat{\Delta}$ in the ensemble set by the cavity $\langle \hat{\Delta}^2\rangle_{\tilde{F}}$ (full line) and in the thermal ensemble at the effective temperature $\langle \hat{\Delta}^2\rangle_{T_{\dot{Q}}} $ (dotted line). In all simulations, we have set $\Gamma = \delta$.  }\label{fig:distributionFunctionFermions}
\end{figure}

For concreteness and simplicity, we restrict the analysis of the present section to a one dimensional system. The kinetic equation derived above, however, is not restricted to one dimension and the qualitative effects of the non-thermal cavity fluctuations are expected to persist in higher dimensional system.

In Fig.~\ref{fig:distributionFunctionFermions}a, the fermionic distribution function $\tilde{F}(\varepsilon)$ obtained from the kinetic equation is shown in black. Additionally, the dotted lines show thermal distribution function for varying temperatures $T$ in the scale of $T_{\rm LF}$. These define an energy-dependent temperature $T(\varepsilon)$ through $\tilde{F}(\varepsilon) = \tanh( \varepsilon / 2 T(\varepsilon))$. Graphically, $T(\varepsilon)$ is the temperature of the thermal distribution which crosses $\tilde{F}(\varepsilon)$ at $\varepsilon$. In this sense, the non-thermal distribution can be interpreted as getting hotter at higher energies.

The distribution function describes how a state at energy $\varepsilon$ is occupied. To completely describe the system, we also need to specify the energies at which there are states. This information is carried by the spectral function (or the density of state). To that end, in Fig.~\ref{fig:distributionFunctionFermions}b-d. we show the density of state, in different cases, superimposed over the distribution function.

At low energies $\varepsilon\ll \delta$, the fermionic distribution function displays thermal-like behavior at the low frequency effective temperature of the boson $T_{\rm LF}$. As such, a fermionic system whose bandwidth is much smaller than the cavity frequency behaves in the normal phase as if it were thermal at the LFET of the cavity. This corresponds to the situation depicted in Fig.~\ref{fig:distributionFunctionFermions}b, as well as the normal phase described in Ref.~\cite{tolle2025fluctuation}. In this regime, the spectral density of the cavity ${\rm Im}~\tilde{\mathcal{D}}^R(\omega)$ can be approximated (at low frequencies) like a linear function, while the Keldysh component satisfies a low-frequency FDT at temperature $T_{\rm LF}$. Then, the cavity propagator enters in the effective fermion action of Eq.~\eqref{eq:switchingToSymmetrizedInteractions} exactly like a bath spectral function. As such, in the low frequency regime, the cavity fluctuations effectively act as an ohmic bath at the temperature $T_{\rm LF}$, and the fermions thermalize to said temperature.

The second case, illustrated by Fig.~\ref{fig:distributionFunctionFermions}c, is that of the bandwidth of the fermionic system being of the same order of the cavity frequency $\delta$. In this regime, there is non negligible spectral weight in frequency ranges where the distribution function shows non-thermal behavior, leading to non-thermal occupations of the state in the system. In particular, the states with energy $\varepsilon\geq \delta/2$ are occupied more than they would in a thermal state. This is understood as the transition between the two bands $\Delta\varepsilon \geq \delta$, so that cavity fluctuations can effectively excite fermions from the lower band to the upper band. However, such processes have a prefactor ${\rm Im}~\tilde{\mathcal{D}}^R(\Delta\varepsilon)$ in the collision integrals Eq.~\eqref{eq:collisionIntegralForm}, and thus become very weak for $\Delta\varepsilon\gg \delta$. As such, they will be dominated by any other source of collisions in the system, as we will discuss more in detail in the next section when treating the interacting problem.

Lastly, in the superradiant phase, the mean-field term $\hat{H}_{\rm MF} = g\phi\Delta$ opens a gap $2g\phi$ between the bands - as per Eq.~\eqref{eq:MF_singlePartcileEnergies} - pushing the states to higher energies which, from previous remarks, are effectively hotter. As the bandwidth of the symmetry-broken bands scales as $4J^2/g\phi$ for $J\ll g\phi$, the frequency-dependent temperature can be approximated to a constant over the bandwidth, so that the system behave as in equilibrium at the temperature $T_{\phi} = T(g\phi)$. From this, we realize that the effective temperature of the system in the superradiant phase grows with $g\phi$.

This heating behavior can be understood in terms of the effective bath played by the cavity fluctuations. In the broken-symmetry phase, the cavity-mediated self-energy only involves transition with energies $\Delta\varepsilon = 2g\phi \pm \frac{8J^2}{g\phi}$. 
Deep in the superradiant phase, the photon propagator entering the effective fermionic action becomes approximately constant over the relevant frequency range. The effective bath therefore approaches the Markovian limit with the jump operator $\hat{L}_{\rm eff} = {\sqrt{\kappa_{\rm eff}/  L^d}}\hat{\Delta}$ where $ {\kappa_{\rm eff}}  = -  g^2 {\rm Im}~\tilde{\mathcal{D}}^R(2g\phi) $. As discussed before, such a Lindblad master equation leads to an infinite temperature steady-state. However, the finite bandwidth of the fermions prevents the Markovian limit from being reached exactly, resulting in a finite effective temperature. Consequently, the effective temperature increases as the system moves deeper into the broken-symmetry phase, where the cavity bath becomes increasingly Markovian.

Finally, Fig.~\ref{fig:distributionFunctionFermions}e displays (in black) the temperature $T_{\dot{Q}}$ obtained, as in Ref.~\cite{tolle2025fluctuation}, by setting the energy current from the fermions to the cavity to zero, in the normal phase while varying $J/\delta$. As the bandwidth grows, the system feels more of the "hot part" of the distribution and the effective temperature grows with $J/\delta$. Then, to illustrate that the state is actually non thermal when $J\sim \delta$, we show that even if the temperature is fixed via a certain criterion, other observables may not match the value of the thermal ensemble. To that end, the fluctuations of $\Delta$ in both ensembles $\langle \Delta^2\rangle_{T_{\dot{Q}}} $, $\langle \Delta^2\rangle_{\tilde{F}} $ are shown (in blue). As we expect, if the low frequency description of the cavity fluctuation holds i.e. when $J\ll \delta$, the thermal ensemble at $T_{\rm LF}$ describes well the system. However, as the approximation breaks the description of the system as thermal is no longer faithful when computing observables other than the one fixed in the definition of the effective temperature ($\dot{Q}$ in the present case) .

\subsection{Results: correction to mean-field phase diagram}

We now turn to the discussion of the cavity order parameter and the correction to the mean-field transition.

With the knowledge of how states are occupied, Eq.~\eqref{eq:expectationValueDeltaFromGK} can then be used to compute $\langle\Delta\rangle$, which has to be solved self-consistently through the Hartree term absorbed in the Hamiltonian. This allows to capture the superradiant phase transition. In the present approach, the cavity fluctuations are considered to only modify the occupations of the fermions but not their spectral properties. Under this approximation, the kinetic equation serves to solve self-consistently the Dyson equation. As such, by simultaneously doing the mean-field calculation for the retarded part and the kinetic equation, we capture the self-consistent effect of the mean-field + fluctuations of the cavity. 

\begin{figure}[t]
\centering
\resizebox{0.95\textwidth}{!}{
\begin{tikzpicture}
\begin{scope}[xshift = 0cm, yshift = 0cm, scale = 1]
\node at (0,0){\includegraphics{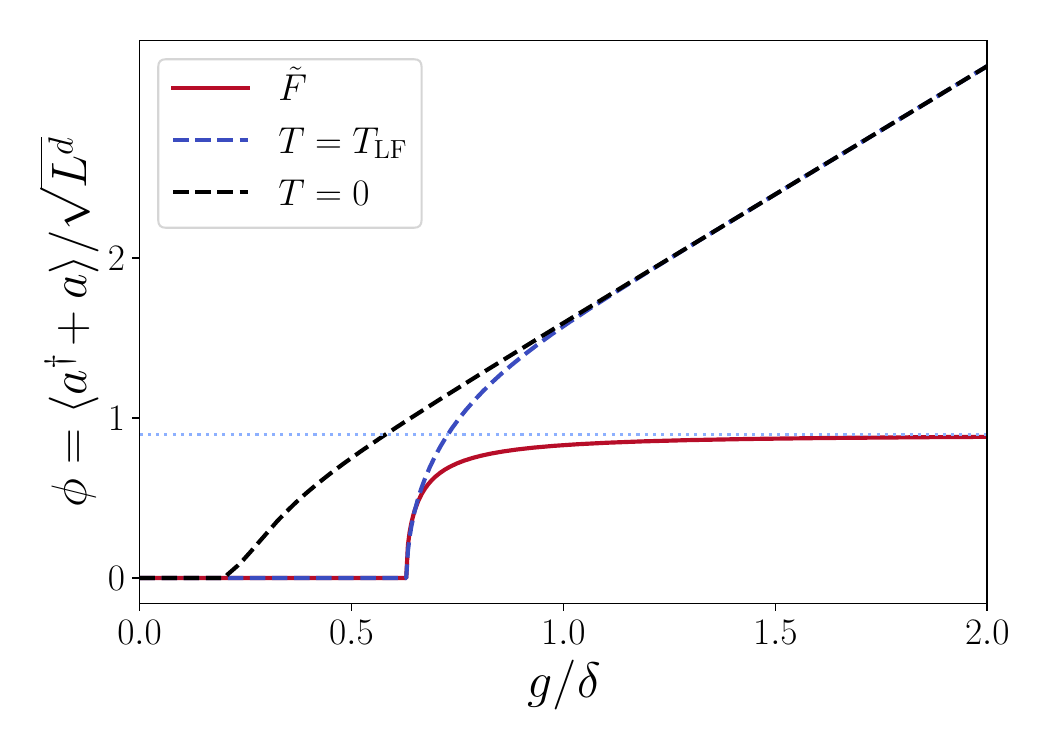}  };
\end{scope}
\begin{scope}[xshift = 17.5cm, yshift = 0cm]
\node at (0,0){\includegraphics[scale = 1]{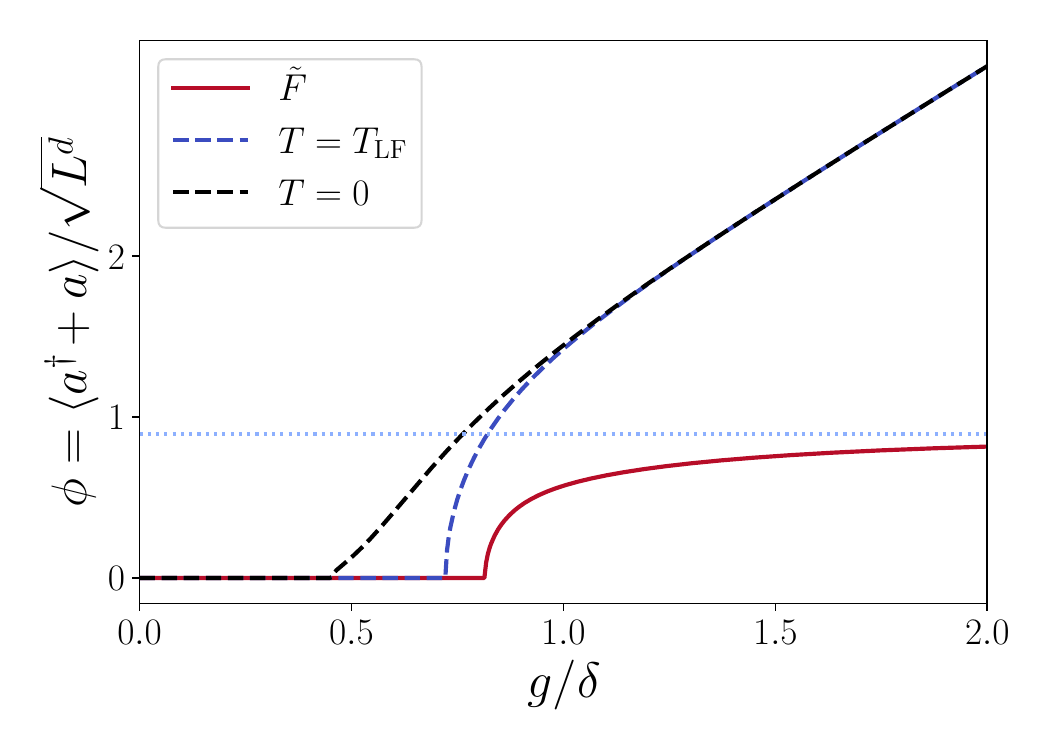}  };
\end{scope}
\node at (-7.5, 6) {\Huge\textbf{a.}};
\node at (10, 6) {\Huge\textbf{b.}};

\end{tikzpicture}
}
\caption{Order parameter obtained from the mean-field + kinetic equation resolution for a low bandwidth $\delta = 10 J$ \textbf{a.} and high bandwidth $\delta = 2 J$ \textbf{b.} fermionic systems. The other parameters are set to $\Gamma = \delta$. 
} \label{fig:MeanFieldResults}
\end{figure}

In Fig.~\ref{fig:MeanFieldResults}, the order parameter resulting from the mean-field + kinetic equation treatment is displayed in both the case of $J\ll\delta$ (panel \textbf{a.}) and $J\sim \delta$ (panel \textbf{b.}). In both figures, the mean-field results of different ensembles -thermal at $T= 0$, thermal at $T = T_{\rm LF}$ and the cavity mediated distribution $\tilde{F}$- are compared.

Overall, the phase transition itself behaves like the finite-temperature transition rather than the ($T=0$) quantum phase transition, as it does in the open Dicke model \cite{torre2013keldysh,kirton2019introduction}. Furthermore, the phase transition itself is not changed and still belongs to the mean-field universality class, as a Hubbard-Stratonovich decoupling shows the saddle point to be exact in the thermodynamic limit.

However, away from the transition and deep into the superradiant phase, the order parameter saturates, contrarily to the equilibrium case. This is understood in terms of the gap opening more and the density of state going to effectively higher temperature regions of the distribution function. As such, as the light-matter coupling $g$ is increased, favoring the formation of a large cavity field, the system heats up which hinders the ordering of the system. The two opposing mechanism lead to the saturation of the order parameter. Furthermore, in the limit of a big gap $g\phi\gg \delta, J$, the distribution function can be expanded at high frequencies in both the equation for the effective temperature and the mean field equation. From this, the asymptotic behavior is captured to be : 
\begin{subequations}
\begin{align}
    \phi &\underset{\scriptstyle g\phi\rightarrow\infty}{\sim} \left( 1 + \frac{\Gamma^2}{4\delta^2}\right)^{-\frac{1}{2}}, \\
    T_{ \phi} &\underset{\scriptstyle g\phi\rightarrow\infty}{\sim} \frac{g^2}{4 T_{\rm LF}} ,
\end{align}
\end{subequations}
notably the temperature scales faster than the gap and the plateau of the order parameter is only set by the losses in units of the cavity frequency. The analytically extracted saturation order parameter is shown by the light blue dotted line in Fig.~\ref{fig:MeanFieldResults}.

Nonetheless, in the regime where the system is non-thermal in the normal phase i.e. $J\sim \delta$, the cavity fluctuations can change the critical coupling. Indeed, in Fig.~\ref{fig:MeanFieldResults}b. , $T_{\rm LF}$ does not describe the system. Moreover, the system takes longer to reach the saturation of the order parameter than the $J\ll \delta$ case. Overall, the larger bandwidth means that the system at a given value of $g\phi$ has states at higher energies, which correspond to higher "temperatures". This means the system takes longer to order, delaying both the symmetry breaking and the saturation.

\subsection{Link Between Kinetic Equation and Heat-Flux}\label{sec:heatFlux}

Having understood the microscopic effects of the dissipative cavity, we now connect this to thermodynamic quantities. We compute the heat-current from the fermions into the cavity in the non-thermal state and find out that the distribution set by the kinetic equation derived from the (self-consistent) Dyson equation exactly cancels out this heat current in the steady state, hereby validating the working hypothesis of Ref.~\cite{tolle2025fluctuation}.

In free-fermions, the energy is a one-body operator and is simply obtained through the equal time Keldysh Green's function : 
\begin{equation}\label{eq:FreeFermionEnergyFromKeldyshGF}
    E(t) = \langle \hat{H}_{\rm FF}\rangle (t) = \frac{\mathrm{i}}{2}\sum_{\mathbf{k}}{\rm tr}~ \check{h}_{\mathbf{k}}\check{G}^K(\mathbf{k} ; t, t ),
\end{equation}
up to a constant. Then, the time variation of energy is expressed in terms of the time derivatives of the Keldysh Green's function. The latter are given by the Dyson equation through the action of the inverse bare propagators $(G^{R/A}_0)^{-1} = \mathrm{i} \partial_t - \check{h}_{\mathbf{k}}$. Following the calculations detailed in Appendix~\ref{app:derivationHeatFlux}, the heat flux is given in the steady state by : 
\begin{equation}\label{eq:HeatFlux}
\dot{Q}=  \frac{2g^2}{L^d} \int \omega \left[ 1 + F_{\Delta}(\omega)  \right] ~  {\rm Im }  \chi^R_{\Delta}(\omega) ~ \left( - {\rm Im }  \mathcal{D}^R(-\omega) \right)  ~ d\omega,
\end{equation}
where $\chi_{\Delta}(\omega)$ is the $\Delta-\Delta$ response function of the fermions and $F_{\Delta}$ is the associated distribution function defined by $\chi_{\Delta}^K(\omega) = F_{\Delta}(\omega) \left[\chi_{\Delta}^R(\omega)  -   \chi_{\Delta}^A(\omega)\right] $. In thermal equilibrium at temperature $T$, it takes the form of a bosonic distribution function  $F_{\Delta}(\omega) = {\rm coth}\frac{\omega}{2T} = 1 + 2n_{\rm B}(\omega)$, where $n_{\rm B}$ is the Bose-Einstein distribution.

Alternatively, the heat-flux can be expressed using the symmetrized boson propagator through (cf. Appendix~\ref{app:derivationHeatFlux}) : 
\begin{equation}\label{eq:HeatFlux_symmetrized}
   \dot{Q} = \frac{2g^2}{L^d} \int   {\omega} \left[ \tilde{B}(\omega) - F_{\Delta}(\omega) \right] ~  {\rm Im }  \chi^R_{\Delta}(\omega)~ {\rm Im} \tilde{\mathcal{D}}^R({\omega}) ~d\omega.
\end{equation}
Under this form the heat flow appears as the frequency-integrated difference between the distribution function of the cavity fluctuations $\tilde{B}$ and order parameter $\Delta$ fluctuations $F_{\Delta}$, weighted by the overlap of their spectral functions.

Both the Kinetic equation and the expression for the heat-flux are derived fully from the Dyson equation, as such we expect them to be equivalent. In the following, we show that the solution of the Kinetic equation in the stationary state indeed nullifies the Heat-flux.

Starting from Eq.~\eqref{eq:SteadyStateHeatFlux_GSigmaExpression}, the heat-flux can be expressed as : 
\begin{align*} 
\dot{Q} &= \frac{1}{2}\int \omega ~{\rm tr}\left[ \check{G}^K \left( \check{\Sigma}^R - \check{\Sigma}^A \right) +\left( \check{G}^A - \check{G}^R \right)  \check{\Sigma}^K  \right]~d\omega d\mathbf{k},
\end{align*}
while the retarded and advance component of the Dyson equation Eq.~\eqref{eq:dysonEquationCheckMatrices_R} give : 
\begin{align*}
    \left[\check{G}^R\right]^{-1}- \left[\check{G}^A\right]^{-1}&=  \check{\Sigma}^A - \check{\Sigma}^R,
\end{align*}
so that in the expression of the heat-flux we write : 
\begin{align*} 
\dot{Q} &=\frac{1}{2}\int \omega ~ {\rm tr}\left[ \check{G}^K \left[\check{G}^A\right]^{-1} - \left[\check{G}^R\right]^{-1}\check{G}^K  +  \check{\Sigma}^K \check{G}^A -  \check{G}^R\check{\Sigma}^K  \right] ~d\omega d\mathbf{k} \\
&=\frac{1}{2}\int \omega ~ {\rm tr}\left[ \left(\left[\check{G}^R\right]^{-1} \check{G}^K \left[\check{G}^A\right]^{-1} - \check{\Sigma}^K \right) \check{G}^A  -\check{G}^R   \left(\left[\check{G}^R\right]^{-1} \check{G}^K \left[\check{G}^A\right]^{-1} - \check{\Sigma}^K \right)  \right] ~d\omega d\mathbf{k}\\
&=\int \omega ~ {\rm tr}\left[ \left(\left[\check{G}^R\right]^{-1} \check{G}^K \left[\check{G}^A\right]^{-1} - \check{\Sigma}^K \right)  \frac{ \check{G}^A  -\check{G}^R   }{2}  \right] ~d\omega d\mathbf{k}. 
\end{align*}
Finally, the distribution function is made explicit through Eq.~\ref{fig:distributionFunctionFermions} : 
\begin{align*}
    \dot{Q} &= \int \omega ~ {\rm tr}\left[ \left( \check{F}\left[\check{G}^A\right]^{-1} - \left[\check{G}^R\right]^{-1} \check{F}- \check{\Sigma}^K \right)  \frac{ \check{G}^A  -\check{G}^R   }{2}  \right] ~d\omega d\mathbf{k} \\
 &= \mathrm{i} \pi \int \omega ~ {\rm tr} \left( \left[\check{F}, \check{h}_{\mathbf{k}} \right]- \check{\Sigma}^K  + \check{\Sigma}^R \check{F} - \check{F}  \check{\Sigma}^A \right)   \check{\mathcal{A}}   ~d\omega d\mathbf{k},
\end{align*}
where we recognize exactly the form of the kinetic equation given in Eq.~\eqref{eq:KineticEquationSteadyState}, while the spectral function $\check{\mathcal{A}}$ and the $\omega$-integration project the equation on-shell. As a consequence of this, a distribution function which solves the on-shell kinetic equation corresponds to a state, whether thermal or not, with exactly vanishing heat-flux between the fermions and the cavity.

\section{Weakly-interacting regime $U\ll J$}\label{sec:finiteU}

The limit of free fermions is very particular in the sense that the quasi-particles have no source of collisions and thus the imaginary part in the inverse propagator is infinitesimal. Due to this infinitesimal collision rate, cavity fluctuations effect always dominate the imaginary part of the Dyson equation and fully set the distribution function no matter how weak is the term. This leads to rather unphysical prediction like the absence of dependence of the distribution on the light-matter coupling $g$ or the very hot high frequency tail $\Delta\varepsilon \gg \delta$ where even far away from the cavity frequency its fluctuations have an effect on the system - which manifests as the collision rate scaling as $1/\Delta\varepsilon^2$.

Away from this limit, the interactions between fermions provide collisions between quasi-particles and ultimately lead to the thermalization of the system. In the present section, the Hubbard interaction are considered to be weak $U\ll  J$ but non zero. In this case, the interplay of the intrinsic thermalization and the non-equilibrium drive by the cavity fluctuations will lead to a crossover from non-thermal to thermal state at a temperature set by the cavity, as well as regularize much of the unphysical results of the non-interacting theory.

The effect of the Hubbard interactions is once again treated through the Dyson equation. We consider that the system remains in the magnetically disordered phase. As such, the Hartree and Fock contributions to the self-energy can be absorbed as a shift of the chemical potential. Furthermore, in the absence of magnetic order, the Green's function of the fermions can be taken as diagonal in spin space
\begin{equation*}
    \check{G}_{\sigma, \sigma^{\prime}}^{R/A/K}(\mathbf{k}, \omega)  = \delta_{\sigma, \sigma^{\prime}}\check{G}^{R/A/K}(\mathbf{k}, \omega),
\end{equation*}
where $\sigma$ is the spin index and $\check{\bullet}$ still denotes a matrix of the sublattice index.

\begin{figure}[htbp]
    \centering
    \resizebox{0.95\textwidth}{!}{%
\begin{tikzpicture}
\begin{scope}[xshift = 0cm, yshift = 0cm, scale = 1]
    \begin{feynman}
\vertex (a) at (0,1) [label = left :{$i ,t, \gamma^c$}];
\vertex (b) at (2,1) [label = right :{$i ,t, \gamma^q$}] ;
  \diagram* {
    (a) -- [dashed, very thick,  edge label = \(U\)]  (b) , 
    };
\vertex (c) at (0,0) [label = left :{$i ,t, \gamma^{\alpha}$}];
\vertex (d) at (2,0) [label = right :{$j ,t^{\prime}\kern-0.2em, \gamma^{\beta}$}] ;
  \diagram* {
    (c) -- [boson,  thick,  edge label = \(\mathcal{D}^{\alpha,\beta}\)]  (d) , 
    };
    \end{feynman}
\end{scope}

\begin{scope}[xshift = 4.25cm, yshift = 0cm, scale = 1.25] 
\begin{feynman}

\vertex (c) at (0,0);
\vertex (d) at (2,0);
\vertex (e) at (0.7,0.8);
\vertex (f) at (1.3,0.8);
\diagram*{
(c) -- [fermion, very thick] (d),
(c) -- [dashed, very thick, bend left = 20] (e),
(d) -- [dashed, very thick, bend right = 20] (f),
(e) -- [fermion, very thick, half left] (f),
(f) -- [fermion, very thick, half left] (e),
};  
\fill (c) circle (2pt);
\fill (d) circle (2pt);
\end{feynman}
\end{scope}

\begin{scope}[xshift =  8cm, yshift = 0cm, scale = 1.25] 
\begin{feynman}
\vertex (a) at (0,0);
\vertex (b) at (1,0);
\vertex (c) at (2,0);
\vertex (d) at (3,0);
\diagram*{
(a) -- [fermion, very thick] (b),
(b) -- [fermion, very thick] (c),
(c) -- [fermion, very thick] (d),
(a) -- [ dashed, very thick, half left] (c),
(b) -- [boson , very thick, half left] (d),
};
\fill (a) circle (2pt);
\fill (b) circle (2pt);
\fill (c) circle (2pt);
\fill (d) circle (2pt);
\end{feynman}
\end{scope}
\node at (-1.5,1.25) {\textbf{a.}};
\node at (-1.5,0.25) {\textbf{b.}};
\node at (4,1.25) {\textbf{c.}};
\node at (7.75,1.25) {\textbf{d.}};
\end{tikzpicture}  
} 
    \caption{\textbf{a.} Hubbard interaction line. \textbf{b.} Cavity-mediated interaction line. \textbf{c.} Second Born self-energy diagram, to which we limit the contributions to the self-energy of the Hubbard interactions. \textbf{d.} Example of a diagram mixing the two types of interaction. All such contributions are neglected. }
    \label{fig:digramsWithHubbard}
\end{figure}
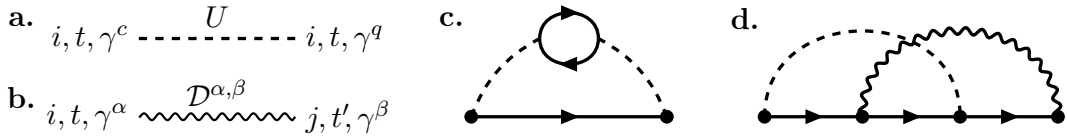

The lowest order diagram relevant to the present study is the second Born contribution shown in Fig.~\ref{fig:digramsWithHubbard}c. We neglect the diagrams involving both cavity-mediated interactions and Hubbard ones like that of Fig.~\ref{fig:digramsWithHubbard}d. This is controlled in the regime where $g^2/L^d \delta \sim U^2/J \ll U$, so that the mixed diagrams are of order $g^2/L^d \delta \times U $ which is negligible with respect to cavity-mediated Fock diagram $\sim g^2/L^d\delta $ and the second Born diagram $\sim U^2/J$. This will turn out to be the regime of the crossover.

The second Born diagram can be expressed as a contribution to the self-energy of the form : 
\begin{subequations}
    \begin{align}
        &\left[\check{\Sigma}_{{\rm hub}}^{K}\right]_{\nu, \nu^{\prime}}(\mathbf{r}, t) = \frac{U^2}{4}\left\{ G^K_{\nu,\nu^{\prime}\kern-0.15em}(\mathbf{r},t)^2 G^K_{\nu^{\prime}\kern-0.25em,\nu}(-\mathbf{r},-t) + 2  G^R_{\nu,\nu^{\prime}\kern-0.15em}(\mathbf{r},t) G^K_{\nu,\nu^{\prime}\kern-0.15em}(\mathbf{r},t)G^A_{\nu^{\prime}\kern-0.25em,\nu}(-\mathbf{r},-t) \right. \\ & + \left. 2  G^A_{\nu,\nu^{\prime}\kern-0.15em}(\mathbf{r},t) G^K_{\nu,\nu^{\prime}\kern-0.15em}(\mathbf{r},t)G^R_{\nu^{\prime}\kern-0.25em,\nu}(-\mathbf{r},-t) + G^R_{\nu,\nu^{\prime}\kern-0.15em}(\mathbf{r},t)^2 G^K_{\nu^{\prime}\kern-0.25em,\nu}(-\mathbf{r},-t)  + G^A_{\nu,\nu^{\prime}\kern-0.15em}(\mathbf{r},t)^2 G^K_{\nu^{\prime}\kern-0.25em,\nu}(-\mathbf{r},-t) 
        \right\}, \notag \\ 
        &\left[\check{\Sigma}_{{\rm hub}}^{R}\right]_{\nu, \nu^{\prime}}(\mathbf{r}, t) = \frac{U^2}{4}\left\{  2  G^R_{\nu,\nu^{\prime}\kern-0.15em}(\mathbf{r},t) G^K_{\nu,\nu^{\prime}\kern-0.15em}(\mathbf{r},t)G^K_{\nu^{\prime}\kern-0.25em,\nu}(-\mathbf{r},-t) +  G^R_{\nu,\nu^{\prime}\kern-0.15em}(\mathbf{r},t)^2 G^A_{\nu^{\prime}\kern-0.25em,\nu}(-\mathbf{r},-t)  \right. \\ 
        & + \left. G^K_{\nu,\nu^{\prime}\kern-0.15em}(\mathbf{r},t)^2 G^A_{\nu^{\prime}\kern-0.25em,\nu}(-\mathbf{r},-t) 
        \right\}, \notag 
    \end{align}
\end{subequations}
where the indices $\nu,\nu^\prime$ are sublattice indices. The self-energy in momentum/frequency space $\Sigma_{\rm hub}^{RAK}(\mathbf{k}, \omega)$ are then obtained through a Fourier transform.

Once again, the Dyson equation can be mapped into a kinetic equation. In the $U\ll J$ regime the quasi-particles are still well defined and upon the projections on the bands the slow dynamics of populations is given by :
\begin{equation}
    \partial_t f_{\mathbf{k}, \nu} = \mathcal{I}_{\mathbf{k}, \nu}^{\rm cav} + \mathcal{I}_{\mathbf{k}, \nu}^{\rm hub},
\end{equation}
where $\mathcal{I}_{\mathbf{k}, \nu}^{\rm cav} $ describes the collisions due to the cavity and is defined in Eq.~\eqref{eq:collisionIntegralForm}, while $\mathcal{I}_{\mathbf{k}, \nu}^{\rm hub}$ comes from the collisions between Hubbard-interacting fermions and is given by :
\begin{align} 
&\mathcal{I}_{\mathbf{k}, \nu}^{\rm hub}= \frac{U^2}{4}\kern-.5em\sum_{\nu_1, \nu_2,\nu_3} \int \frac{ d^d \mathbf{k}^{\prime}d^d \mathbf{q}}{(2\pi)^{2d} } ~ \mathcal{M}^{\nu, \nu_1, \nu_2, \nu_3}_{\mathbf{k}, \mathbf{k^{\prime}}, \mathbf{k+q}, \mathbf{k^{\prime}+q}}~\delta\left( \varepsilon_{\mathbf{k}, \nu}+ \varepsilon_{\mathbf{k^{\prime} + q}, \nu_3} - \varepsilon_{\mathbf{k^{\prime}}, \nu_1} - \varepsilon_{\mathbf{k+q}, \nu_2} \right)\notag\\
&\times \left\{ \left[f_{\mathbf{k+q}, \nu_2}  -f_{\mathbf{k},\nu}\right] \left[ f_{\mathbf{k^{\prime} + q}, \nu_3} f_{\mathbf{k^{\prime}}, \nu_1} - 1  \right] -\left[f_{\mathbf{k+q},\nu_2}f_{\mathbf{k}, \nu}-1\right]  \left[ f_{\mathbf{k^{\prime} + q}, \nu_3} - f_{\mathbf{k^{\prime}}, \nu_1} \right] \right\}, \label{eq:HubbardCollisionIntegral}
\end{align}
where the tensor $\mathcal{M}^{\nu, \nu_1, \nu_2, \nu_3}_{\mathbf{k}, \mathbf{k_1}, \mathbf{k_2}, \mathbf{k_3}}$ quantifies the overlap of the different states via colliding : 
\begin{equation*} \label{eq:Mtensor}
\mathcal{M}^{\nu, \nu_1, \nu_2, \nu_3}_{\mathbf{k}, \mathbf{k_1}, \mathbf{k_2}, \mathbf{k_3}}  ={\rm tr}\left[ \left( \check{P}_{\mathbf{k_1}, \nu_1} \ast \check{P}_{\mathbf{k_2},\nu_2} \right) \times \left( \check{P}_{\mathbf{k_3}, \nu_3} \ast \check{P}_{\mathbf{k}, \nu} \right) \right],
\end{equation*}
where $\ast$ is the Hadamard element-wise product defined by $\left(\check{A}\ast\check{B}\right)_{i,j} = \left(\check{A}\right)_{i,j} \left(\check{B}\right)_{i,j}$.

\begin{figure}[t]
\centering
\includegraphics[width=10cm,height=6cm]{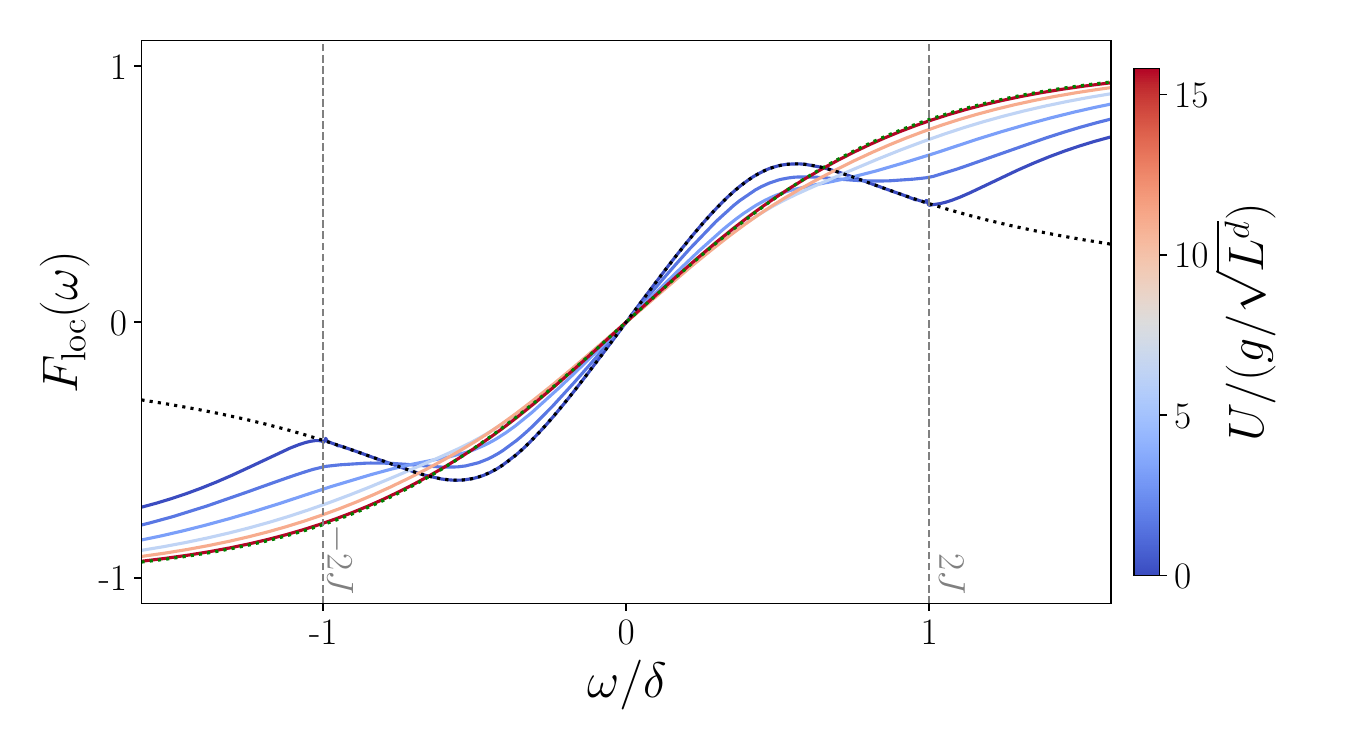}
\caption{The figure shows the distribution function $F_{\rm loc}(\omega)$ across the crossover from the cavity-mediated at low interactions (blue) to the thermal state at high interactions (red). The analytical solution in the free fermions limit (U=0) is shown by the black dotted line, while the green one displays the fit of the high U result by a thermal distribution function. The gray dotted lines signal the bounds of the fermions bandwidth beyond which the spectral function is only made of the tails of the lorentzian of the quasi-particle peaks. There, the quasi-particle approximation fails, explaining the disparity between the $U= 0 $ numerical solution and the analytical solution, however the interactions lead to thermalization even outside the quasi-particle region. The simulations were done for $J = \delta/2$ and $\Gamma =\delta$. }\label{fig:crossover}
\end{figure}
Most importantly, the energy conserving Dirac delta and trigonometric formulas allow to show \cite{kamenevFieldTheoryNonEquilibrium2011} that $f_{\mathbf{k}, \nu} = {\rm tanh} \frac{\varepsilon_{\mathbf{k}, \nu}}{2 T }$ is a solution of $\mathcal{I}^{\rm hub}[f] = 0 $ for all temperatures $T$. That is, the collisions between interacting fermions draws the system toward a thermal state but does not fix the temperature of system.

As such, in the regime $g^2/\delta L^d\ll U^2/J$ where fermion-fermions collisions dominate fermion-photon collisions, we expect that the system settles into a thermal steady state. However, since the fermion-fermion collisions cannot fix the temperature, we expect the temperature to be set by the cavity fluctuations. The approach of Ref.~\cite{tolle2025fluctuation} appears applicable in this regime, as the condition of vanishing heat-flux is both physically justified and explicitly supported by the case of free fermions detailed in Sec.~\ref{sec:heatFlux}.

Indeed, numerical resolution of the self-consistent Dyson equation based on Ref.~\cite{picano_quantum_2021,schuler_nessi_2020} shows a crossover from the non-thermal steady-state to a thermal state when $g^2/\delta L^d\sim U^2/J$,or equivalently $g/\sqrt{L^d}\sim U$ since $J\sim \delta$ is required to even have a non-thermal state. In Fig.~\ref{fig:crossover}, the local distribution function $F_{\rm loc}(\omega) = {\rm Im}~{\rm tr}\check{G}^K_{\rm loc}(\omega) / 2{\rm Im}~{\rm tr}\check{G}^R_{\rm loc}(\omega)$, where $\check{G}^{R/A/K}_{\rm loc}(\omega) = \int_{\rm BZ} \check{G}^{R/A/K}(\mathbf{k}, \omega)d\mathbf{k}$, is displayed for different values of the Hubbard collision rate $U^2/J$ relatively to the cavity-mediated collision rate $g^2/\delta L^d$. Justification of the distribution function as $F_{\rm loc}(\omega)$ as well as details of the numerics are relegated to Appendix \ref{App:LocalApprox}.

The regime where the non-thermal effect may survive is still within reach. If we take $J\sim \delta$, which we have seen to be necessary to have an effectively non-thermal state, and small interactions of the order of $U\sim 0.1 J$ and light matter couplings $g\sim  \delta$, then the non-thermal state survives to system sizes of $L^d\sim 10^2$.

However, in both the thermal and non-thermal regimes, the collisions rates are small and the effect can be cut-off by other sources of collisions not considered here. In particular, Eq.~\eqref{eq:collisionIntegralForm} highlights that deep in the superradiant phase the cavity-mediated collision rate diminishes, so that the high-$g$ behavior of the theory eventually fails. Even when the Hubbard collisions are not small and the system can be considered to be in a thermal state, at some point the temperature will no longer be set by the cavity but by other baths/noise sources. As such, care has to be taken in the study of the superradiant phase and other sources of noise may be more relevant than the cavity fluctuations.

Overall, the cavity fluctuations play the role of a frequency-dependent injection/extraction of energy into the fermions, while the (mean-field) cavity-mediated interactions change the spectrum of the fermions in the superradiant phase and thus how the cavity fluctuations inject energy into the system. Such instances of a frequency-dependent injection of energy and interaction are well known to give rise to bistability. The simplest example being a non-linear oscillator driven coherently and with losses which then displays bistability as a function of the driving strength. We attribute the results of Ref.~\cite{tolle2025fluctuation} to this phenomenology. In the superradiant phase, as we have seen off-resonant cavity fluctuations heat up the system. However, in the Mott insulator phase of the Hubbard model, when the staggered potential is such that transition between the doublon of a sublattice and the singlon of the other becomes resonant with the cavity, the system is cooled. This gives rise to the second branch of solutions. Such phenomena do not appear in the low-$U$ limit as the system does not have sharp resonances but the excitations in $\chi_{ \Delta}$ form a continuum instead.

\section{Conclusions}\label{sec:conclusions}

In this work we have studied a Fermi-Hubbard model coupled to a driven-dissipative cavity mode. The problem has recently received significant attention due to the interplay between local many-body interactions and cavity-mediated ones, leading to a rich phase diagram where fluctuations play an important role~\cite{bezvershenko2021dicke,tolle2025fluctuation,steadystate2026diagram}.

Using Keldysh diagrammatics we have derived and solved the quantum kinetic equation for the fermionic degrees of freedom, integrating out the cavity mode which leads to a source of non-equilibrium noise and dissipation for the fermions. In absence of Hubbard interactions we have first shown that the light-matter coupling leads in general to a non-equilibrium distribution for the matter degrees of freedom. This is ultimately due to the fact that a single mode cavity does not allow sufficient momentum scattering to truly thermalize the system. Depending on the parameters we have highlighted regimes, notably the low-bandwidth one, where one can identify  a low-frequency effective temperature for the fermions and show to coincide with the one of the cavity mode. Furthermore we have shown the role of fluctuations in controlling the mean-field phase diagram and the transition to the superradiant phase. Finally we have demonstrated explicitly that the non-thermal steady-state obtained for the fermions corresponds to a vanishing heat current between matter and cavity system, a working hypothesis of Ref.~\cite{tolle2025fluctuation} which we have derived here microscopically from the kinetic equation.

We have then considered the role of fermionic (Hubbard) interactions in the weak-coupling regime, by adding a fermionic self-energy diagram corresponding to the second-order Born approximation to the cavity-mediated one. We have shown that genuine many-body interactions lead to a crossover from non-thermal to full thermal distribution.

Our work highlights the role of non-equilibrium effects in driven cavity-quantum matter problems. In the future it would be interesting to extend our analysis beyond the steady-state and to focus on the full out of equilibrium dynamics. To this extent it would be natural to use methods developed to solve efficiently the time evolution of correlated electrons systems such as out of equilibrium extensions of Dynamical Mean-Field Theory~\cite{RevModPhys.86.779,picano_quantum_2021,picano2025quantum,picano2025heatingdynamicscorrelatedfermions}. Furthermore,  our Keldysh diagrammatic approach can be naturally extended beyond lattice models to explore non-equilibrium effects in cold fermions coupled
to cavities~\cite{zwettler2025nonequilibrium,marijanovic_quench_2026}.

\section*{Acknowledgements}
We acknowledge useful discussions with Giuliano Chiriaco, Gian-Marcello Andolina, and thank Antonio Picano for valuable discussions and assistance with the numerical implementation.

\paragraph{Funding information}
This work has received funding from the European Research Council (ERC) under the European Union's Horizon 2020 research and innovation program (Grant agreement No. 101002955 -- CONQUER).
The  computations  were  performed  on  the  Coll\`ege de France IPH computer cluster. 

\paragraph*{Code availability} The code used to obtain the results presented in this work is available at Ref.~\cite{code_repository}.

\begin{appendix}

\section{Derivation of the cavity-mediated collision integral}\label{app:DerivationOfKineticEquation}

For the purposes of the calculation at hand, we introduce the orthonormal basis $(\vec{u}_{\mathbf{k}}, \vec{v}_{\mathbf{k}}, \vec{w}_{\mathbf{k}})$ so that the single-particle Hamiltonian defined in Eq.~\eqref{eq:defh_k} is along $\mathbbm{1}$ and $ \vec{u}_{\mathbf{k}} \cdot \vec{\check{\tau}}$ i.e. 
\begin{equation}
  \check{h}_{{\mathbf{k}}} = \bar{\varepsilon}_{\mathbf{k}} \mathbbm{1} + \Delta\varepsilon_{\mathbf{k}} \vec{u}_{\mathbf{k}} \cdot \vec{\check{\tau}},
\end{equation} 
where $\vec{\check{\tau}}$ are the Pauli matrices, and from Eq.~\eqref{eq:defh_k} we get $u_{\mathbf{k}, x} = \frac{J   ( 1 + \cos k_{\parallel} )}{\sqrt{2J^2 (1+ \cos k_{\parallel}) + g^2 \phi^2} }$, $u_{\mathbf{k}, y} = \frac{ J \sin k_{\parallel} }{\sqrt{2J^2 (1+ \cos k_{\parallel}) + g^2 \phi^2} }$
, and $u_{\mathbf{k}, z} = \frac{ g\phi}{\sqrt{2J^2 (1+ \cos k_{\parallel}) + g^2 \phi^2} }$.

Furthermore, up to a rotation of the basis, it can be supposed that : 
\begin{equation} \label{eq:uTau_zrotation}
\tau_z( \vec{u}_{\mathbf{k}} \cdot \vec{\check{\tau}}  ) \tau_z = \alpha_{\mathbf{k}} \vec{u}_{\mathbf{k}} \cdot \vec{\check{\tau}}  + \beta_{\mathbf{k}} \vec{v}_{\mathbf{k}} \cdot \vec{\check{\tau}}  ,
\end{equation}
where $\alpha_{\mathbf{k}} = 2 u_{\mathbf{k}, z}^2  -1  $ and $\beta_{\mathbf{k}} = \sqrt{1 - \alpha_{\mathbf{k}} }$. In particular, in the normal phase $\alpha_{\mathbf{k}} = -1 $.

The projector can be expressed as $ \check{P}_{\mathbf{k}, \nu} = \frac{\vec{u}_{\mathbf{k}}\cdot \vec{\check{\tau}} + \nu }{2{\nu}}  $, so that Eq.~\eqref{eq:onShellCommutingAnsatz} becomes : 
\begin{equation} \label{eq:onShellCommutingAnsatz_appendix}
\check{G}^{K}(\mathbf{k}, \omega) = - \mathrm{i}2\pi\sum_{\nu = \pm} f_{\mathbf{k}, \nu}~ \delta(\omega - \varepsilon_{\mathbf{k}, \nu} )~ \frac{\vec{u}_{\mathbf{k}}\cdot \vec{\check{\tau}} + \nu }{2{\nu}} ,
\end{equation} 
Within this form, the collision integral appearing in the right hand side of the kinetic equation simply becomes :
\begin{equation*}
    \mathcal{I}_{\mathbf{k}, \nu} = {\Sigma}^K_{\nu}(\mathbf{k}, \omega = \varepsilon_{\mathbf{k}, \nu}) - \left\{ \Sigma^R_{\nu}(\mathbf{k}, \omega = \varepsilon_{\mathbf{k}, \nu})  - \Sigma^A_{\nu}(\mathbf{k}, \omega = \varepsilon_{\mathbf{k}, \nu}) \right\} f_{\mathbf{k}, \nu}, 
\end{equation*}
where ${\Sigma}_{\nu}^{R/A/K}(\mathbf{k}, \omega ) = {\rm tr}~ \check{P}_{\mathbf{k}, \nu} \check{\Sigma}_{\nu}^{R/A/K}(\mathbf{k}, \omega)$, and the frequency has been fixed on-shell by the Dirac delta and a frequency integration in Eq.~\eqref{eq:dysonEquationCheckMatrices_K}.

For simplicity's sake, in the rest of the intermediary calculation the $\mathbf{k}$ indices are implicit. This doesn't pose any issue as the cavity mediated interaction conserve the momentum of the fermions.

The first term in the collision integral is ${\Sigma}^K_{\nu}(\mathbf{k}, \omega)$ which can be expressed by projecting Eq.~\eqref{eq:self-energyGeneralExpression_detailK_tau} on the eigenstates of $\check{h}$ : 
\begin{align*}
    {\Sigma}^K_{\nu}(\omega) &= \frac{g^2}{L^d}\sum_{\nu^{\prime}}\int_{\tilde{\omega}} {\rm Im}~\tilde{\mathcal{D}}^{R}(\tilde{\omega}) \delta\kern-0.1em\left( \omega - \tilde{\omega} -  \varepsilon_{\nu^{\prime}}\right)  \left\{  \tilde{B}(\tilde{\omega})    f_{\nu^{\prime}}   + 1 \right\} ~  {\rm tr}  \frac{\vec{u}\cdot \vec{\check{\tau}} + \nu}{2{\nu}}  \check{\tau}^{z}~ \frac{\vec{u}_{\mathbf{k}}\cdot \vec{\check{\tau}} + \nu^{\prime}  }{2{\nu^{\prime} }} \check{\tau}^{z}, 
\end{align*}
then, using Eq.~\eqref{eq:uTau_zrotation} and the fact that $\vec{u}\perp\vec{v}$, the overlap of the two matrices can be computed 
$ {\rm tr}  \frac{\vec{u}\cdot \vec{\check{\tau}} + \nu}{2{\nu}}  \check{\tau}^{z}~ \frac{\vec{u}_{\mathbf{k}}\cdot \vec{\check{\tau}} + \nu^{\prime}  }{2{\nu^{\prime} }} \check{\tau}^{z} = \frac{\alpha + \nu\nu^{\prime}}{2\nu\nu^{\prime}},$
so that : 
\begin{align*}
    {\Sigma}^K_{\nu}(\omega) &= \frac{g^2}{L^d} \sum_{\nu^{\prime}}\int d\tilde{\omega}~ {\rm Im}~\tilde{\mathcal{D}}^{R}( \omega-  \varepsilon_{\nu^{\prime}})   \left\{  \tilde{B}( \omega-  \varepsilon_{\nu^{\prime}})    f_{\nu^{\prime}}   + 1 \right\} \frac{\alpha + \nu\nu^{\prime}}{2\nu\nu^{\prime}}. 
\end{align*}
On the other hand, the second term in the collision integral is given by :
\begin{align*}
    \Sigma^R_{\nu}(\mathbf{k}, \omega)  - \Sigma^A_{\nu}(\mathbf{k}, \omega) &=\frac{g^2}{L^d}\sum_{\nu^{\prime}} {\rm Im}~\tilde{\mathcal{D}}^{R}(\omega - \varepsilon_{\nu^{\prime}})\left[  \tilde{B}(\omega - \varepsilon_{\nu^{\prime}})   +f_{\nu^{\prime}}   \right]  \frac{\alpha + \nu\nu^{\prime}}{2\nu\nu^{\prime}}, 
\end{align*}
so that finally the collision appears as : 
\begin{align*}
    \mathcal{I}_{\nu}& = \frac{g^2}{ L^d}\sum_{\nu^{\prime}} {\rm Im}~\tilde{\mathcal{D}}^{R}(\varepsilon_{\nu} - \varepsilon_{\nu^{\prime}})\left\{  \tilde{B}( \varepsilon_{\nu}-  \varepsilon_{\nu^{\prime}})    f_{\nu^{\prime}}   + 1   -  \left(  \tilde{B}(\varepsilon_{\nu} - \varepsilon_{\nu^{\prime}})   +f_{\nu^{\prime}}   \right)f_{\nu} \right\}  \frac{\alpha + \nu\nu^{\prime}}{2\nu\nu^{\prime}},\\ 
    & = \frac{g^2}{L^d}\sum_{\nu^{\prime}} {\rm Im}~\tilde{\mathcal{D}}^{R}(\varepsilon_{\nu} - \varepsilon_{\nu^{\prime}})\left\{  \tilde{B}( \varepsilon_{\nu}-  \varepsilon_{\nu^{\prime}})   ( f_{\nu^{\prime}}  -f_{\nu}) + 1   -  f_{\nu^{\prime}}f_{\nu}   \right\}  \frac{\alpha + \nu\nu^{\prime}}{2\nu\nu^{\prime}}. 
\end{align*}
Now, from Eq.~\eqref{eq:RAK_symmetrizedPhotonPropagator}, we have ${\rm Im}~\tilde{\mathcal{D}}^{R}(\omega \rightarrow 0 ) = 0 $, while  $\tilde{B}\times{\rm Im}~\tilde{\mathcal{D}}^{R}(\omega \rightarrow 0 )  =  - \frac{\Gamma/2}{(\Gamma/2)^2 +\delta^{2} }\neq 0 $, so that the term corresponding to $\nu^{\prime} = \nu$ in the sum above all but the terms containing $\tilde{B}$ drop out. Thus, the term is simply proportional to $f_\nu - f_\nu$ and vanishes. As such, only the $\nu^{\prime} = \bar{\nu}$ term contributes to the collision integral, leaving us with : 
\begin{align*}
    \mathcal{I}_{\nu} =\frac{g^2}{L^d} {\rm Im}~\tilde{\mathcal{D}}^{R}(\varepsilon_{\nu} - \varepsilon_{\bar{\nu}})\left\{  \tilde{B}( \varepsilon_{\nu}-  \varepsilon_{\bar{\nu}})   ( f_{\bar{\nu}}  -f_{\nu}) + 1   -  f_{\bar{\nu}}f_{\nu}   \right\}  \frac{1 - \alpha }{2}, 
\end{align*}
restoring the $\mathbf{k}$ index and realizing that $\varepsilon_{\nu} - \varepsilon_{\bar{\nu}} = \nu \Delta \varepsilon$ and $\alpha = 2 u_{z}^2 - 1$, we are left with : 
\begin{equation}\label{eq:collisionIntegralForm_appendix}
   \kern-0.175em \mathcal{I}_{\mathbf{k}, \nu}^{\rm cav}\kern-0.1em = \kern-0.1em\frac{g^2}{L^d} \kern-0.1em\left(1 \kern-0.1em- \kern-0.1em u_{\mathbf{k}, z}^2\right) \kern-0.1em\nu {\rm Im}\tilde{\mathcal{D}}^{R}(\Delta \varepsilon_{\mathbf{k}}) \left\{ \nu\tilde{B}( \Delta \varepsilon_{\mathbf{k}})    f_{\bar{\nu}} (\mathbf{k}) \kern-0.1em +\kern-0.1em 1  \kern-0.1em-\kern-0.1em \left( \nu  \tilde{B}(\Delta \varepsilon_{\mathbf{k}})  \kern-0.1em + \kern-0.1em f_{\bar{\nu}}(\mathbf{k}) \kern-0.1em  \right)\kern-0.1em  f_{\nu}(\mathbf{k})\right\}\kern-0.1em  . 
\end{equation}

\section{Derivation of the heat-flow in the U=0 limit}\label{app:derivationHeatFlux}

In this section, we derive the heat-flux from the fermions to the cavity in the non-thermal steady state, in the case of $U = 0$. To compute this (dynamic) quantity, the assumption of time-translation invariant Green's functions has to be relaxed $\check{G}^{a, b}(t, t^{\prime}) \neq \check{G}^{a, b}(t - t^{\prime}) $, and the product of kernels no longer simplifies to a product in frequency representation. In this paragraph, $\circ$ explicitly denotes the product of two kernels as operators $(A\circ B) (t, t^{\prime}) \equiv \int A(t, z) B(z, t^{\prime})~dz$, similarly to a continuously-indexed matrix product.

In this more general case, the Keldysh component of the 
\begin{equation}\label{eq:FullTimeDysonEquation}
\left(\mathrm{i} \partial_t - \check{h}_{\mathbf{k}} - \hat{\Sigma}^R\right) \circ \hat{G}^K - \hat{\Sigma}^K \circ \hat{G}^A = 0 ,
\end{equation}
where we have used that the inverse bare propagator can be expressed as $(\check{G}_{0}^{R})^{-1} = \mathrm{i} \partial_t - \check{h}_{\mathbf{k}}$. Then, using the fact that $\mathrm{i}\partial_t \circ \check{G} = \mathrm{i}\partial_t\check{G}$ and $\check{G}\circ\mathrm{i} \partial_t = \mathrm{i}\partial_{t^{\prime}}\check{G}$, Eq.~\eqref{eq:FullTimeDysonEquation} and its (Hermitian) conjugate can be written as : 
\begin{subequations}\label{eq:KeldyshDysonEquationTimeDepence_detailDeriative}
    \begin{align}
\mathrm{i} \partial_t  \check{G}^K  &= \check{h}_{\mathbf{k}} \check{G}^K + \check{\Sigma}^R \circ \check{G}^K  + \check{\Sigma}^K \circ \check{G}^A,\\
  \mathrm{i} \partial_{t^{\prime}}  \check{G}^K  &=  -  \check{G}^K  \check{h}_{\mathbf{k}}  -  \check{G}^K \circ \check{\Sigma}^A  - \check{G}^R \circ \check{\Sigma}^K  .
    \end{align}
\end{subequations}

Then, the heat-flux is the time-derivative of the fermions energy as given in Eq.~\eqref{eq:FreeFermionEnergyFromKeldyshGF} which is expressed as : 
\begin{align*}
    \dot{Q} &= \partial_t \langle\hat{H}_{\rm ff}\rangle = \frac{\mathrm{i}}{2} \sum_{\mathbf{k}} {\rm tr} ~ \check{h}_{\mathbf{k}} \partial_t  \check{G}^K({\mathbf{k}} ; t,t) + \check{h}_{\mathbf{k}}  \partial_{t^{\prime}}  \check{G}^K({\mathbf{k}} ; t,t) , \\
    &= \frac{1}{2} \sum_{\mathbf{k}} {\rm tr}  \left[ \check{G}^K({\mathbf{k}}; t,t) , \check{h}^2_{\mathbf{k}} \right] +    \check{h}_{\mathbf{k}} \kern-0.2em \left(
    \check{\Sigma}^R \kern-0.2em \circ \check{G}^K \kern-0.25em -  \check{G}^K \kern-0.2em \circ \check{\Sigma}^A \kern-0.25em   +  \check{\Sigma}^K \kern-0.2em \circ \check{G}^A \kern-0.25em  - \check{G}^R \circ \check{\Sigma}^K \right)\kern-0.2em ({\mathbf{k}}; t,t) ,  \\
&=\frac{1}{2} \sum_{\mathbf{k}}{\rm tr} \left( \check{\Sigma}^R \circ \check{G}^K \check{h}_{\mathbf{k}} - \check{h}_{\mathbf{k}} \check{G}^K \circ \check{\Sigma}^A + \check{\Sigma}^K \circ \check{G}^A \check{h}_{\mathbf{k}} - \check{h}_{\mathbf{k}}\check{G}^R  \kern-0.2em  \circ \check{\Sigma}^K \right)({\mathbf{k}}; t,t),
\end{align*}
where in the last line we have the cyclicity of the trace to cancel the trace of a commutator as well as always put the $\check{h}$ against the green's function rather than the self-energy. This is so $\check{h}$ can be eliminated from the expression by using Eq.~\eqref{eq:KeldyshDysonEquationTimeDepence_detailDeriative} the other way around, yielding after cancellations :
\begin{align*}
\dot{Q} &= -\frac{\mathrm{i}}{2} \sum_{\mathbf{k}}{\rm tr} \left( \check{\Sigma}^R  \circ \partial_{t^{\prime}} \check{G}^K + \partial_t \check{G}^K \circ\check{\Sigma}^A +  \check{\Sigma}^K \circ \partial_{t^{\prime}}\check{G}^A +  \partial_t  \check{G}^R \circ \check{\Sigma}^K \right)({\mathbf{k}}; t,t) , 
\end{align*}
Then, going to the Wigner representation of two point functions, whereby any function $F(t, t^{\prime})$ has a unique Wigner transform $F_{W}(T, \omega)$ from which it can be reconstructed via :  
\begin{align*}
    F(t, t^{\prime}) = \int_{\omega}  F_{W}\left(\frac{t + t^{\prime}}{2}, \omega\right) e^{-\mathrm{i} \omega (t - t^{\prime})} ,
\end{align*}
where $\int_{\omega} = \int \frac{d\omega}{2\pi}$.

The time derivatives act on the Wigner representation as $\partial_{t} \longmapsto \frac{1}{2}\partial_T - \mathrm{i} \omega$ and $\partial_{t^{\prime}} \longmapsto \frac{1}{2}\partial_T + \mathrm{i} \omega$, while the equal-time evaluation gives an integration on $\omega$. As such, the heat-flux reads : 
\begin{align*} 
\dot{Q} &= -\frac{\mathrm{i}}{4} \sum_{\mathbf{k}} \int_{\omega} {\rm tr}  \left( \check{\Sigma}^R  \circ \partial_{T} \check{G}^K + \partial_T \check{G}^K \circ\check{\Sigma}^A +  \check{\Sigma}^K \circ \partial_{T}\check{G}^A +  \partial_T \check{G}^R \circ \check{\Sigma}^K \right)\\
&-\frac{1}{2} \sum_{\mathbf{k}} \int_{\omega} {\rm tr}  \left( \check{\Sigma}^R  \circ (\omega \check{G}^K) -  (\omega\check{G}^K) \circ\check{\Sigma}^A +  \check{\Sigma}^K \circ (\omega\check{G}^A) -   (\omega\check{G}^R) \circ \check{\Sigma}^K \right),
\end{align*}
where the convolution $\circ$ maps onto the Moyal product in the Wigner representation. In the steady state, the $T$-derivatives vanish while the Moyal product reduces to a product in frequency, so that : 
\begin{equation}\label{eq:SteadyStateHeatFlux_GSigmaExpression}
    \dot{Q}_{\rm SS} = -  \frac{1}{2} \sum_{\mathbf{k}} \int_{\omega} \omega~ {\rm tr} \left( \check{\Sigma}^R \check{G}^K \kern-0.25em - \check{\Sigma}^A\check{G}^K\kern-0.25em + \check{G}^R\check{\Sigma}^K \kern-0.25em- \check{\Sigma}^K\check{G}^A \right). 
\end{equation}
Upon further inspection of the Keldysh indices in Eq.~\eqref{eq:SteadyStateHeatFlux_GSigmaExpression}, the expression for the heat-flux can be expressed as : 
\begin{equation*} \label{eq:intermediateHeatFlux_Diagrams_CommutatorForm}
\dot{Q}_{\rm SS} = - \frac{1}{2}  \sum_{\mathbf{k}} \int_{\omega} \omega ~ \check{\rm tr} \left[ \hat{\Sigma} ~, ~\hat{ G}_0\right]^{1,2}_{\rm k.i.} ,
\end{equation*}
where $\left[ ~, ~\right]_{\rm k.i.}$ is the commutator taken on the Keldysh indices, ${\bullet}^{1, 2}$ refers to the Keldysh component of said commutator while $\check{\rm tr}$ signifies the trace on the sublattice indices. Now, the causal structure, imposes that the Green's function and the self-energy are upper diagonal, thus their product and their commutator also are. In particular, $[ \hat{\Sigma} ~, ~\hat{ G}_0]^{2, 1}_{\rm k.i.} = 0$ and thus : 
\begin{align*}
    \left[ \hat{\Sigma} ~, ~\hat{ G}\right]^{1, 2}_{\rm k.i.} &= \sum_{a}\left[ \hat{\Sigma} ~, ~\hat{ G}\right]^{a, \bar{a}}_{\rm k.i.} = {\rm tr}_{\rm k.i.} \left[ \hat{\Sigma} ~, ~\hat{ G}\right]_{\rm k.i.} \gamma^q, 
\end{align*}
and, since the full trace can be written as ${\rm tr} = \check{\rm tr}~ {\rm tr}_{\rm k.i.}$, in the heat-flux : 
\begin{align*}
\dot{Q}_{\rm SS}& =  -  \frac{1}{2}  \sum_{\mathbf{k}} \int_{\omega} \omega ~ {\rm tr}~ \gamma^q \left[ \hat{\Sigma} ~, ~\hat{ G}\right]_{\rm k.i.} =  -  \frac{1}{2}  \sum_{\mathbf{k}} \int_{\omega} \omega ~ {\rm tr}~ \left(\gamma^q \hat{\Sigma} - \hat{\Sigma} \gamma^q\right)\hat{ G}  . 
\end{align*}
Now, the self-energy in Eq.~\eqref{eq:FockSelfEnergy_KeldyshIndices}, can be written with implicit Keldysh indices as : 
\begin{equation*}
      \hat{\Sigma}(\mathbf{k}, \omega) = \mathrm{i} \frac{g^2}{L^d} \sum_{\alpha, \beta}\int_{\tilde{\omega}} \tilde{\mathcal{D}}^{\alpha, \beta}(\tilde{\omega}) \gamma^{\alpha} ~\check{\tau}^{z} ~\hat{G}(\mathbf{k}, \omega - \tilde{\omega}) ~\gamma^{\beta}  ~ \check{\tau}^{z}. 
\end{equation*}
Then, realizing that $\gamma^{\alpha} \gamma^q = \gamma^q\gamma^\alpha = \gamma^{\bar{\alpha}}$, the heat-flux can be written : 
\begin{align*}
    \dot{Q}_{\rm SS}& =- \mathrm{i} \frac{g^2}{2 L^d} \sum_{\mathbf{k}, \alpha,\beta} \int_{\omega, \tilde{\omega}}\kern-0.5em \omega ~  \tilde{\mathcal{D}}^{\alpha, \beta}(\tilde{\omega}) ~\left\{   {\rm tr} ~ \hat{G}(\mathbf{k}, \omega)  \gamma^{\bar{\alpha}} \check{\tau}^{z} \hat{G}(\mathbf{k}, \omega - \tilde{\omega}) \gamma^{\beta}   \check{\tau}^{z} \right. \\ &-\left.{\rm tr} ~ \hat{G}(\mathbf{k}, \omega) \gamma^{\alpha} \check{\tau}^{z} \hat{G}(\mathbf{k}, \omega - \tilde{\omega}) \gamma^{\bar{\beta}}   \check{\tau}^{z} \right\}, 
\end{align*}
the second term is then subject to the changes of variable $\omega \mapsto \omega + \tilde{\omega}$ , $\tilde{\omega} \mapsto -\tilde{\omega}$ and $\alpha\longleftrightarrow\beta$, and using the cyclicity of the trace the term inside the trace is the same in both term , leaving us with 
\begin{align*}
    \dot{Q}_{\rm SS}& = -  \mathrm{i} \frac{g^2}{2 L^d} \sum_{\mathbf{k}, \alpha,\beta} \int_{\omega, \tilde{\omega}} \left\{  \omega ~  \tilde{\mathcal{D}}^{\alpha, \beta}(\tilde{\omega})  -( \omega - \tilde{\omega}) ~  \tilde{\mathcal{D}}^{ \beta, \alpha}(-\tilde{\omega})\right\} {\rm tr} ~ \hat{G}(\mathbf{k}, \omega)    \gamma^{\bar{\alpha}} \check{\tau}^{z} \hat{G}(\mathbf{k}, \omega - \tilde{\omega}) \gamma^{\beta}   \check{\tau}^{z} . 
\end{align*}
Now, the symmetrized propagator defined in Eq.~\eqref{eq:RAK_symmetrizedPhotonPropagator} satisfies the symmetry $\tilde{\mathcal{D}}^{ \beta, \alpha}(-\tilde{\omega})= \tilde{\mathcal{D}}^{\alpha, \beta}(\tilde{\omega}) $, so that the term proportional to $\omega$ drops out and the $\omega$-integration can be carried. One recognizes the response function : 
\begin{equation}\label{eq:definitionResponseFunction}
    \chi^{\alpha,\beta}_{\Delta}(\tilde{\omega} ) = -\frac{\mathrm{i} }{2} \sum_{\mathbf{k}} \int_\omega {\rm tr} ~\hat{G}(\mathbf{k}, \omega)  \gamma^{\alpha}\check{\tau}^{z} \hat{G}(\mathbf{k}, \omega - \tilde{\omega}) \gamma^{\beta}   \check{\tau}^{z}, 
\end{equation}
so that the heat-flux reads :
\begin{equation}\label{eq:HeatFlux_KeldyshChiAndD}
      \dot{Q}_{\rm SS} =  \frac{g^2}{ L^d} \sum_{\alpha,\beta} \int_{\omega} \omega~ \tilde{\mathcal{D}}^{\alpha, \beta}({\omega})   \chi^{\bar{\alpha}, \beta}_{\Delta}(\omega).  
\end{equation}
Now, in order to obtain an expression similar to that of Ref.~\cite{tolle2025fluctuation}, the Keldysh indices need to be made explicit and the symmetrized photon propagator has to be reverted to the natural/non-symmetrized one. 
Developing the sum on $\alpha$ and $\beta$, remembering that $\chi$ has the structure of an inverse bosonic propagator, gives : 
\begin{equation} \label{eq:Heat-FluxalphaBetaSum_develloped}
\dot{Q}=  \frac{g^2}{L^d} \int   {\omega} \left(\tilde{\mathcal{D}}^R({\omega}) ~ \chi_{\Delta}^K({\omega}) + \tilde{\mathcal{D}}^K({\omega}) ~ \chi_{\Delta}^A({\omega}) \right)  ~ d{\omega}.
\end{equation} 
From Eq.~\eqref{eq:definitionResponseFunction}, one easily shows that $\chi^{\alpha, \beta}_{\Delta}(\omega) = \chi^{\beta, \alpha}_{\Delta}(-\omega)$. Using this and the definition of the symmetrized photon propagator, one can write : 
\begin{align*} 
\int \omega \tilde{\mathcal{D}}^R (\omega) \chi^K_{\Delta}(\omega) ~d\omega &= \frac{1}{2}  \int \omega \left[ {\mathcal{D}}^R(\omega) +  {\mathcal{D}}^A(-\omega)\right] \chi^K_{\Delta}(\omega) ~d\omega \\
&=  \frac{1}{2}  \int \omega \left[ {\mathcal{D}}^R(\omega) -  {\mathcal{D}}^A(\omega)\right] \chi^K_{\Delta}(\omega) ~d\omega\\
\int \omega \tilde{\mathcal{D}}^K (\omega) \chi^A_{\Delta}(\omega) ~d\omega &= \frac{1}{2}  \int \omega\left[ \mathcal{D}^K (\omega) + \mathcal{D}^K (-\omega)\right] \chi^A_{\Delta}(\omega)~d\omega \\
&=  - \frac{1}{2}  \int \omega \mathcal{D}^K (\omega) \left[ \chi^R_{\Delta}(\omega) - \chi^A_{\Delta}(\omega)\right] ~ d\omega.
\end{align*}
Then, one utilizes the Fluctuation-Dissipation Relations : 
\begin{align*} 
 \mathcal{D}^K (\omega)   &= 2 \mathrm{i} ~ {\rm Im }  \mathcal{D}^R(\omega) \\
  \chi^K_{\Delta} (\omega)  &= 2\mathrm{i}  F_{\Delta}(\omega) ~  {\rm Im }  \chi^R_{\Delta}(\omega),
\end{align*}
where $F_{\Delta}(\omega)$ is the bosonic-like occupation function of the $\Delta$ collective excitations. With that, one writes : 
\begin{equation*}
\dot{Q}=  \frac{2g^2}{L^d} \int \omega \left[ 1-  F_{\Delta}(\omega)    \right] ~  {\rm Im }  \chi^R_{\Delta}(\omega) ~\left( - {\rm Im }  \mathcal{D}^R(\omega) \right) ~ d\omega.
\end{equation*}
Finally, one performs the change of variable $\omega \longmapsto -\omega$, and using the symmetry of the response function, obtain : 
\begin{equation} \label{eq:HeatFluxDistributionFunction}
\dot{Q}=  \frac{2g^2}{L^d} \int \omega \left[ 1 + F_{\Delta}(\omega)  \right] ~  {\rm Im }  \chi^R_{\Delta}(\omega) ~  \left( - {\rm Im }  \mathcal{D}^R(-\omega) \right) ~ d\omega. 
\end{equation}
Now, for a thermal state, the distribution function of the order parameter fluctuations $ 1+ F_{\Delta}(\omega) = 1+ {\rm coth}(\beta \omega /2) = 2 / (1 - e^{-\beta\omega})$ so that we recover the result of Ref.~\cite{tolle2025fluctuation} : 
\begin{align*}
\dot{Q}=  \frac{4g^2}{ L^d} \int \frac{\omega}{1 - e^{-\beta\omega}}  ~  {\rm Im }  \chi^R_{\Delta}(\omega) ~ \frac{\Gamma/2}{(\omega + \delta)^2 + (\Gamma / 2)^2} ~ d\omega.
\end{align*}

Using the symmetry of $\tilde{D}$ and $\chi_{\Delta}$, Eq.~\eqref{eq:Heat-FluxalphaBetaSum_develloped} can be turned into : 
\begin{align}
\dot{Q}&=  \frac{g^2}{2L^d} \int   {\omega} \left\{ \left[\tilde{\mathcal{D}}^R({\omega}) -  \tilde{\mathcal{D}}^A({\omega}) \right] ~ \chi_{\Delta}^K({\omega}) + \tilde{\mathcal{D}}^K({\omega}) ~ \left[\chi_{\Delta}^A({\omega}) - \chi_{\Delta}^R({\omega})\right]\right\}  ~ d{\omega}\notag \\
&= \frac{2g^2}{L^d} \int   {\omega} \left[ \tilde{B}(\omega) - F_{\Delta}(\omega) \right] ~ {\rm Im} \tilde{\mathcal{D}}^R({\omega}) ~  {\rm Im }  \chi^R_{\Delta}(\omega)~d\omega,  \label{eq:symmetrizedHeatFlowFormula}
\end{align}
where the integrand is now even with respect to frequency and thus the region around $\omega = \delta$ is explicitly as important as the one around $\omega = - \delta$. This form shows the heat-flow as a distance between the distribution function $\tilde{B}(\omega) $ and $ F_{\Delta}(\omega)$.

\section{Local approximation to the Hubbard self-energy}\label{App:LocalApprox}

In order to reduce the numerical complexity of the implementation of the second Born self-energy, we perform the so-called Local approximation to the self-energy whereby we neglect the $k$-dependence of the Hubbard self-energy : 
\begin{equation}
    \check{\Sigma}_{\rm hub}(\omega) \simeq \check{\Sigma}_{\rm loc, hub}(\omega) =  \check{\Sigma}_{\rm hub}(\textbf{r} = 0, \omega) = \frac{1}{L^d}\sum_{k}  \check{\Sigma}_{\rm hub}(\textbf{k}, \omega), 
\end{equation}
and the $k$ dependence is implied by the energy dependence though the dispersion relation. The local approximation, often used in DMFT, is known to still thermalize the system \cite{eckstein_thermalization_2009}.

Owing to the locality of the Hubbard interactions, the local self-energy only depends on the local Greens function :
\begin{equation}
    \check{G}_{\rm loc}(\omega) =  \check{G}(\textbf{r} = 0, \omega) = \frac{1}{L^d}\sum_{k}  \check{G}(\textbf{k}, \omega),
\end{equation}
through :
\begin{align}
    &\left[\check{\Sigma}_{\rm  loc, hub}^{K}\right]_{\kern-0.1em \mu, \nu }\kern-0.25em( t) \kern-0.1em= \kern-0.1em \frac{U^2}{4}\kern-0.1em\left\{ \kern-0.25em\left[\check{G}^K_{\rm loc}\right]_{\kern-0.1em \mu,\nu }\kern-0.25em( t)^2 \kern-0.25em\left[\check{G}^K_{\rm loc}\right]_{\kern-0.1em \nu ,\mu}\kern-0.25em( -t) \kern-0.25em+ \kern-0.25em2  \kern-0.25em\left[\check{G}^R_{\rm loc}\right]_{\kern-0.1em \mu,\nu }\kern-0.25em( t) \kern-0.25em\left[\check{G}^K_{\rm loc}\right]_{\kern-0.1em \mu,\nu }\kern-0.25em( t)\kern-0.25em\left[\check{G}^A_{\rm loc}\right]_{\kern-0.1em \nu,\mu}\kern-0.25em( -t)\kern-0.1em \right. \notag\\ & + \left. \kern-0.25em 2 \kern-0.25em\left[\check{G}^A_{\rm loc}\right]_{\kern-0.1em \mu,\nu } \kern-0.25em\left[\check{G}^K_{\rm loc}\right]_{\kern-0.1em \mu,\nu }\kern-0.25em(t)\kern-0.25em\left[\check{G}^R_{\rm loc}\right]_{\kern-0.1em \nu ,\mu}\kern-0.25em( -t)\kern-0.25em + \kern-0.25em\left[\check{G}^R_{\rm loc}\right]_{\kern-0.1em \mu,\nu }\kern-0.25em( t)^2 \kern-0.25em\left[\check{G}^K_{\rm loc}\right]_{\kern-0.1em \nu,\mu}\kern-0.25em( -t)\kern-0.25em  + \kern-0.25em\left[\check{G}^A_{\rm loc}\right]_{\kern-0.1em \mu,\nu }\kern-0.25em( t)^2 \kern-0.25em\left[\check{G}^K_{\rm loc}\right]_{\kern-0.1em \nu,\mu}\kern-0.25em( -t) 
    \right\}, \notag \\ 
    & \left[\check{\Sigma}_{{\rm loc, hub}}^{R}\right]_{\mu, \nu }\kern-0.25em(t) \kern-0.1em= \kern-0.1em \frac{U^2}{4}\left\{  2  \kern-0.25em\left[\check{G}^R_{\rm loc}\right]_{\mu,\nu }\kern-0.25em{\textstyle( t)} \kern-0.25em\left[\check{G}^K_{\rm loc}\right]_{\mu,\nu }\kern-0.25em( t)\kern-0.25em\left[\check{G}^K_{\rm loc}\right]_{\nu ,\mu}\kern-0.25em( -t) +  \kern-0.25em\left[\check{G}^R_{\rm loc}\right]_{\mu,\nu }\kern-0.25em(t)^2 \kern-0.25em\left[\check{G}^A_{\rm loc}\right]_{\nu ,\mu}\kern-0.25em(-t)  \right.\notag \\ 
    & + \left. \kern-0.25em\left[\check{G}^K_{\rm loc}\right]_{\mu,\nu }\kern-0.25em( t)^2 \kern-0.25em\left[\check{G}^A_{\rm loc}\right]_{\nu ,\mu}\kern-0.25em(-t) 
    \right\}. \label{eq:localHubbardSelfEnergy}
\end{align}
It can be noted that, even in the quasi particle limit, the local Green's function is a smooth function of frequency, contrarily to Green's functions resolved in $k$ which have sharp features making numerical integration complicated.

On the other, the cavity-mediated self-energy should not be treated with the local approximation. Physically, the cavity-mediated interactions are infinite range, oppositely to the local Hubbard self-energy, and thus poorly captured by local-self energies. Mathematically, because ${\rm Im}\mathcal{D^R}(\omega)$ is broad in frequencies the $k$ summation of the local approximation averages over a lot of different $k$ at a fixed frequency.

Instead, since we are in the regime where $g/\sqrt{L^d}\ll J$ and $U\ll J$, we expect the spectral function to still have very sharp quasi-particles. Having neglected all real part of self-energy, the spectral function can be approximated to :
\begin{align*}
    \check{G}^{R}(\mathbf{k}, \omega) - \check{G}^{A}(\mathbf{k}, \omega) &= - \mathrm{i}2\pi\sum_{\nu = \pm}  \delta(\omega - \varepsilon_{\mathbf{k}, \nu} )~ \frac{\vec{u}_{\mathbf{k}}\cdot \vec{\check{\tau}} + \nu }{2{\nu}} ,
\end{align*}
while, provided the occupation of the mode $k$ is only dependent on its energy $\varepsilon_k$, the Keldysh component of the Green's function is approximated : 
\begin{equation}\label{eq:DefinitionDistributionFunction}
    \check{G}^{K}(\mathbf{k}, \omega) = - \mathrm{i}2\pi\sum_{\nu = \pm} F( \varepsilon_{\mathbf{k}, \nu})~ \delta(\omega - \varepsilon_{\mathbf{k}, \nu} )~ \frac{\vec{u}_{\mathbf{k}}\cdot \vec{\check{\tau}} + \nu }{2{\nu}} .
\end{equation}
Within those approximation the cavity-mediated self-energy are accessible analytically from Eq~\eqref{eq:self-energyGeneralExpression_detail_tau} with their expression involving the photon propagator and the occupation function.

The numerics would easily be adapted to keep the real part of the Hubbard self-energy. In the low $U$, regime it would mostly lead to a shift in the energies $\varepsilon_{\mathbf{k}}\longmapsto \tilde{\varepsilon_{\mathbf{k}}}$ which would have to be determined by solving $\tilde{\varepsilon_{\mathbf{k}}} = {\varepsilon_{\mathbf{k}}} + {\rm Re}\Sigma_{\rm  loc, hub}(\tilde{\varepsilon_{\mathbf{k}}})$. In the present study, the most physically relevant effect of the Hubbard interactions is the thermalization and thus those energy shifts are ignored. While the real part of the self-energy is also responsible for the loss of quasi-particle weight, those effects remain small in the regime of $U\ll J$.

Finally, the distribution function is extracted from the local Green's function by realizing that : 
\begin{align*}
    {\rm tr}~ \check{G}^{K}_{\rm loc}(\omega) &= \frac{1}{L^d} \sum_{k} {\rm tr} ~\check{G}^{K}(k, \omega) \\
&= - \mathrm{i}2\pi\sum_{\nu = \pm} \frac{1}{L^d} \sum_{k}  F( \varepsilon_{\mathbf{k}, \nu})~ \delta(\omega - \varepsilon_{\mathbf{k}, \nu} )  \underbrace{{\rm tr}  \frac{\vec{u}_{\mathbf{k}}\cdot \vec{\check{\tau}} + \nu }{2{\nu}} }_{=1}\\
&= - \mathrm{i}2\pi \frac{1}{L^d} \sum_{\substack{k \\ \nu = \pm}}   F(\omega)~ \delta(\omega - \varepsilon_{\mathbf{k}, \nu} ) \\
&=   F(\omega) ~  \frac{- \mathrm{i}2\pi}{L^d} \sum_{\substack{k \\ \nu = \pm}}  \delta(\omega - \varepsilon_{\mathbf{k}, \nu} ) \\
&=   F(\omega) ~ 2\mathrm{i} {\rm Im} ~{\rm tr}~ \check{G}^{R}_{{\rm loc}}( \omega),
\end{align*}
so that : 
\begin{equation}
     F(\omega) = \frac{  {\rm Im} ~{\rm tr}~ \check{G}^{K}_{{\rm loc}}( \omega)  }{  2{\rm Im} ~{\rm tr}~ \check{G}^{R}_{{\rm loc}}( \omega)  } \equiv F_{\rm loc}(\omega). 
\end{equation}
This, also justifies why in the quasi-particle approximation the distribution function $F(\omega) $ defined in Eq.~\eqref{eq:DefinitionDistributionFunction} coincides with the local distribution function $F_{\rm loc}(\omega)$. Anyway, away from the quasi-particle limit, the $k$-resolved spectral function can have broad features from the incoherent spectral weight. There the distribution function $F(\mathbf{k},\omega)$ can no longer be approximated to be frequency independent $F(\mathbf{k}, \omega) = F_\mathbf{k}$ and the collision integral approach breaks down.

\bibliography{biblio}

@misc{code_repository,
  note   = {\url{ https://github.com/julessueir/FermionsOpenCavity}}
}

@article{raimond2001manipulating,
  title = {Manipulating quantum entanglement with atoms and photons in a cavity},
  author = {Raimond, J. M. and Brune, M. and Haroche, S.},
  journal = {Rev. Mod. Phys.},
  volume = {73},
  issue = {3},
  pages = {565--582},
  numpages = {0},
  year = {2001},
  month = {Aug},
  publisher = {American Physical Society},
  doi = {10.1103/RevModPhys.73.565},
  url = {https://link.aps.org/doi/10.1103/RevModPhys.73.565}
}

@article{ritsch2012cold,
  title = {Cold atoms in cavity-generated dynamical optical potentials},
  author = {Ritsch, Helmut and Domokos, Peter and Brennecke, Ferdinand and Esslinger, Tilman},
  journal = {Rev. Mod. Phys.},
  volume = {85},
  issue = {2},
  pages = {553--601},
  numpages = {0},
  year = {2013},
  month = {Apr},
  publisher = {American Physical Society},
  doi = {10.1103/RevModPhys.85.553},
  url = {https://link.aps.org/doi/10.1103/RevModPhys.85.553}
}

@article{Mivehvar02012021,
author = {Farokh Mivehvar and Francesco Piazza and Tobias Donner and Helmut Ritsch},
title = {Cavity QED with quantum gases: new paradigms in many-body physics},
journal = {Advances in Physics},
volume = {70},
number = {1},
pages = {1--153},
year = {2021},
publisher = {Taylor \& Francis},
doi = {10.1080/00018732.2021.1969727},
URL = {https://doi.org/10.1080/00018732.2021.1969727},
}

@book{kamenevFieldTheoryNonEquilibrium2011,
  title = {Field {{Theory}} of {{Non-Equilibrium Systems}}},
  author = {Kamenev, Alex},
  date = {2011-09-08},
  year= {2011},
  edition = {1},
  publisher = {Cambridge University Press},
  doi = {10.1017/CBO9781139003667},
  url = {https://www.cambridge.org/core/product/identifier/9781139003667/type/book},
  urldate = {2025-06-20},
  isbn = {978-0-521-76082-9 978-1-139-00366-7}
}

@article{thompsonFieldTheoryManybody2023,
  title = {Field Theory of Many-Body {{Lindbladian}} Dynamics},
  author = {Thompson, Foster and Kamenev, Alex},
  date = {2023-08},
  year= {2023},
  journaltitle = {Annals of Physics},
  shortjournal = {Annals of Physics},
  volume = {455},
  pages = {169385},
  issn = {00034916},
  doi = {10.1016/j.aop.2023.169385},
  url = {https://linkinghub.elsevier.com/retrieve/pii/S0003491623001719},
  urldate = {2025-07-02},
  langid = {english}
}

@article{siebererKeldyshFieldTheory2016,
  title = {Keldysh Field Theory for Driven Open Quantum Systems},
  author = {Sieberer, L M and Buchhold, M and Diehl, S},
  date = {2016-09-01},
  year = {2016},
  journaltitle = {Reports on Progress in Physics},
  shortjournal = {Rep. Prog. Phys.},
  volume = {79},
  number = {9},
  pages = {096001},
  issn = {0034-4885, 1361-6633},
  doi = {10.1088/0034-4885/79/9/096001},
  url = {https://iopscience.iop.org/article/10.1088/0034-4885/79/9/096001},
  urldate = {2025-06-20}
}

@article{sieberer2025universality,
  title = {Universality in driven open quantum matter},
  author = {Sieberer, Lukas M. and Buchhold, Michael and Marino, Jamir and Diehl, Sebastian},
  journal = {Rev. Mod. Phys.},
  volume = {97},
  issue = {2},
  pages = {025004},
  numpages = {66},
  year = {2025},
  month = {Jun},
  publisher = {American Physical Society},
  doi = {10.1103/RevModPhys.97.025004},
  url = {https://link.aps.org/doi/10.1103/RevModPhys.97.025004}
}

@article{fazio2025manybody,
	title = {Many-body open quantum systems},
	pages = {99},
	author = {Fazio, Rosario and Keeling, Jonathan and Mazza, Leonardo and Schirò, Marco},
	journal = {SciPost Phys. Lect. Notes},
	year = {2025},
	publisher = {SciPost},
	doi = {10.21468/SciPostPhysLectNotes.99},
	url = {https://scipost.org/10.21468/SciPostPhysLectNotes.99}
}

@article{dimer2007proposed,
  title = {Proposed realization of the Dicke-model quantum phase transition in an optical cavity QED system},
  author = {Dimer, F. and Estienne, B. and Parkins, A. S. and Carmichael, H. J.},
  journal = {Phys. Rev. A},
  volume = {75},
  issue = {1},
  pages = {013804},
  numpages = {14},
  year = {2007},
  month = {Jan},
  publisher = {American Physical Society},
  doi = {10.1103/PhysRevA.75.013804},
  url = {https://link.aps.org/doi/10.1103/PhysRevA.75.013804}
}

@article{torre2013keldysh,
  title = {Keldysh approach for nonequilibrium phase transitions in quantum optics: Beyond the Dicke model in optical cavities},
  author = {Torre, Emanuele G. Dalla and Diehl, Sebastian and Lukin, Mikhail D. and Sachdev, Subir and Strack, Philipp},
  journal = {Phys. Rev. A},
  volume = {87},
  issue = {2},
  pages = {023831},
  numpages = {20},
  year = {2013},
  month = {Feb},
  publisher = {American Physical Society},
  doi = {10.1103/PhysRevA.87.023831},
  url = {https://link.aps.org/doi/10.1103/PhysRevA.87.023831}
}

@article{cabllero2015quantum,
  title = {Quantum Optical Lattices for Emergent Many-Body Phases of Ultracold Atoms},
  author = {Caballero-Benitez, Santiago F. and Mekhov, Igor B.},
  journal = {Phys. Rev. Lett.},
  volume = {115},
  issue = {24},
  pages = {243604},
  numpages = {6},
  year = {2015},
  month = {Dec},
  publisher = {American Physical Society},
  doi = {10.1103/PhysRevLett.115.243604},
  url = {https://link.aps.org/doi/10.1103/PhysRevLett.115.243604}
}

@article{larson2008cold,
  title = {Cold Fermi atomic gases in a pumped optical resonator},
  author = {Larson, Jonas and Morigi, Giovanna and Lewenstein, Maciej},
  journal = {Phys. Rev. A},
  volume = {78},
  issue = {2},
  pages = {023815},
  numpages = {8},
  year = {2008},
  month = {Aug},
  publisher = {American Physical Society},
  doi = {10.1103/PhysRevA.78.023815},
  url = {https://link.aps.org/doi/10.1103/PhysRevA.78.023815}
}

@article{piazza2014quantum,
  title = {Quantum kinetics of ultracold fermions coupled to an optical resonator},
  author = {Piazza, Francesco and Strack, Philipp},
  journal = {Phys. Rev. A},
  volume = {90},
  issue = {4},
  pages = {043823},
  numpages = {19},
  year = {2014},
  month = {Oct},
  publisher = {American Physical Society},
  doi = {10.1103/PhysRevA.90.043823},
  url = {https://link.aps.org/doi/10.1103/PhysRevA.90.043823}
}

@article{chen2014superradiance,
  title = {Superradiance of Degenerate Fermi Gases in a Cavity},
  author = {Chen, Yu and Yu, Zhenhua and Zhai, Hui},
  journal = {Phys. Rev. Lett.},
  volume = {112},
  issue = {14},
  pages = {143004},
  numpages = {5},
  year = {2014},
  month = {Apr},
  publisher = {American Physical Society},
  doi = {10.1103/PhysRevLett.112.143004},
  url = {https://link.aps.org/doi/10.1103/PhysRevLett.112.143004}
}

@article{keeling2014fermionic,
  title = {Fermionic Superradiance in a Transversely Pumped Optical Cavity},
  author = {Keeling, J. and Bhaseen, M. J. and Simons, B. D.},
  journal = {Phys. Rev. Lett.},
  volume = {112},
  issue = {14},
  pages = {143002},
  numpages = {5},
  year = {2014},
  month = {Apr},
  publisher = {American Physical Society},
  doi = {10.1103/PhysRevLett.112.143002},
  url = {https://link.aps.org/doi/10.1103/PhysRevLett.112.143002}
}

@article{mivehvar2017superradiant,
  title = {Superradiant Topological Peierls Insulator inside an Optical Cavity},
  author = {Mivehvar, Farokh and Ritsch, Helmut and Piazza, Francesco},
  journal = {Phys. Rev. Lett.},
  volume = {118},
  issue = {7},
  pages = {073602},
  numpages = {6},
  year = {2017},
  month = {Feb},
  publisher = {American Physical Society},
  doi = {10.1103/PhysRevLett.118.073602},
  url = {https://link.aps.org/doi/10.1103/PhysRevLett.118.073602}
}

@article{kollath2016ultracold,
  title = {Ultracold Fermions in a Cavity-Induced Artificial Magnetic Field},
  author = {Kollath, Corinna and Sheikhan, Ameneh and Wolff, Stefan and Brennecke, Ferdinand},
  journal = {Phys. Rev. Lett.},
  volume = {116},
  issue = {6},
  pages = {060401},
  numpages = {5},
  year = {2016},
  month = {Feb},
  publisher = {American Physical Society},
  doi = {10.1103/PhysRevLett.116.060401},
  url = {https://link.aps.org/doi/10.1103/PhysRevLett.116.060401}
}

@article{Sheikhan2016cavity,
  title = {Cavity-induced chiral states of fermionic quantum gases},
  author = {Sheikhan, Ameneh and Brennecke, Ferdinand and Kollath, Corinna},
  journal = {Phys. Rev. A},
  volume = {93},
  issue = {4},
  pages = {043609},
  numpages = {15},
  year = {2016},
  month = {Apr},
  publisher = {American Physical Society},
  doi = {10.1103/PhysRevA.93.043609},
  url = {https://link.aps.org/doi/10.1103/PhysRevA.93.043609}
}

@article{piazza2014umklapp,
  title = {Umklapp Superradiance with a Collisionless Quantum Degenerate Fermi Gas},
  author = {Piazza, Francesco and Strack, Philipp},
  journal = {Phys. Rev. Lett.},
  volume = {112},
  issue = {14},
  pages = {143003},
  numpages = {5},
  year = {2014},
  month = {Apr},
  publisher = {American Physical Society},
  doi = {10.1103/PhysRevLett.112.143003},
  url = {https://link.aps.org/doi/10.1103/PhysRevLett.112.143003}
}

@article{bezvershenko2021dicke,
  title = {Dicke Transition in Open Many-Body Systems Determined by Fluctuation Effects},
  author = {Bezvershenko, Alla V. and Halati, Catalin-Mihai and Sheikhan, Ameneh and Kollath, Corinna and Rosch, Achim},
  journal = {Phys. Rev. Lett.},
  volume = {127},
  issue = {17},
  pages = {173606},
  numpages = {6},
  year = {2021},
  month = {Oct},
  publisher = {American Physical Society},
  doi = {10.1103/PhysRevLett.127.173606},
  url = {https://link.aps.org/doi/10.1103/PhysRevLett.127.173606}
}

@article{kirton2019introduction,
author = {Kirton, Peter and Roses, Mor M. and Keeling, Jonathan and Dalla Torre, Emanuele G.},
title = {Introduction to the Dicke Model: From Equilibrium to Nonequilibrium, and Vice Versa},
journal = {Advanced Quantum Technologies},
volume = {2},
number = {1-2},
pages = {1800043},
doi = {https://doi.org/10.1002/qute.201800043},
url = {https://advanced.onlinelibrary.wiley.com/doi/abs/10.1002/qute.201800043},
year = {2019}
}

@article{orso2025self,
  title = {Self-Organized Cavity Bosons beyond the Adiabatic Elimination Approximation},
  author = {Orso, Giuliano and Zakrzewski, Jakub and Deuar, Piotr},
  journal = {Phys. Rev. Lett.},
  volume = {134},
  issue = {18},
  pages = {183405},
  numpages = {7},
  year = {2025},
  month = {May},
  publisher = {American Physical Society},
  doi = {10.1103/PhysRevLett.134.183405},
  url = {https://link.aps.org/doi/10.1103/PhysRevLett.134.183405}
}

@article{halati2025from,
  title = {From Light-Cone to Supersonic Propagation of Correlations by Competing Short- and Long-Range Couplings},
  author = {Halati, Catalin-Mihai and Sheikhan, Ameneh and Morigi, Giovanna and Kollath, Corinna and J\"ager, Simon B.},
  journal = {Phys. Rev. Lett.},
  volume = {135},
  issue = {19},
  pages = {190402},
  numpages = {8},
  year = {2025},
  month = {Nov},
  publisher = {American Physical Society},
  doi = {10.1103/tt11-vcpr},
  url = {https://link.aps.org/doi/10.1103/tt11-vcpr}
}

@article{tolle2025fluctuation,
  title = {Fluctuation-Induced Bistability of Fermionic Atoms Coupled to a Dissipative Cavity},
  author = {Tolle, Luisa and Sheikhan, Ameneh and Giamarchi, Thierry and Kollath, Corinna and Halati, Catalin-Mihai},
  journal = {Phys. Rev. Lett.},
  volume = {134},
  issue = {13},
  pages = {133602},
  numpages = {8},
  year = {2025},
  month = {Mar},
  publisher = {American Physical Society},
  doi = {10.1103/PhysRevLett.134.133602},
  url = {https://link.aps.org/doi/10.1103/PhysRevLett.134.133602}
}

@article{steadystate2026diagram,
  title = {Steady-state diagram of interacting fermionic atoms coupled to dissipative cavities},
  author = {Tolle, Luisa and Sheikhan, Ameneh and Giamarchi, Thierry and Kollath, Corinna and Halati, Catalin-Mihai},
  journal = {Phys. Rev. Res.},
  volume = {8},
  issue = {2},
  pages = {023041},
  numpages = {27},
  year = {2026},
  month = {Apr},
  publisher = {American Physical Society},
  doi = {10.1103/318z-q249},
  url = {https://link.aps.org/doi/10.1103/318z-q249}
}

@article{chiriaco2022critical,
  title = {Critical light-matter entanglement at cavity mediated phase transitions},
  author = {Chiriac\`o, Giuliano and Dalmonte, Marcello and Chanda, Titas},
  journal = {Phys. Rev. B},
  volume = {106},
  issue = {15},
  pages = {155113},
  numpages = {9},
  year = {2022},
  month = {Oct},
  publisher = {American Physical Society},
  doi = {10.1103/PhysRevB.106.155113},
  url = {https://link.aps.org/doi/10.1103/PhysRevB.106.155113}
}

@article{ortu2026pauli,
  title = {Pauli Crystal Superradiance},
  author = {Ortu\~no-Gonzalez, Daniel and Lin, Rui and Stefaniak, Justyna and Baumg\"artner, Alexander and Natale, Gabriele and Donner, Tobias and Chitra, R.},
  journal = {Phys. Rev. Lett.},
  volume = {136},
  issue = {8},
  pages = {083405},
  numpages = {7},
  year = {2026},
  month = {Feb},
  publisher = {American Physical Society},
  doi = {10.1103/g5fp-ws7z},
  url = {https://link.aps.org/doi/10.1103/g5fp-ws7z}
}

@article{dmytruk2021gauge,
  title = {Gauge fixing for strongly correlated electrons coupled to quantum light},
  author = {Dmytruk, Olesia and Schir\'o, Marco},
  journal = {Phys. Rev. B},
  volume = {103},
  issue = {7},
  pages = {075131},
  numpages = {18},
  year = {2021},
  month = {Feb},
  publisher = {American Physical Society},
  doi = {10.1103/PhysRevB.103.075131},
  url = {https://link.aps.org/doi/10.1103/PhysRevB.103.075131}
}

@article{schwlawin2022cavity,
    author = {Schlawin, F. and Kennes, D. M. and Sentef, M. A.},
    title = {Cavity quantum materials},
    journal = {Applied Physics Reviews},
    volume = {9},
    number = {1},
    pages = {011312},
    year = {2022},
    month = {02},
    issn = {1931-9401},
    doi = {10.1063/5.0083825},
    url = {https://doi.org/10.1063/5.0083825},
}

@article{baumann2010dicke,
	Author = {Baumann, Kristian and Guerlin, Christine and Brennecke, Ferdinand and Esslinger, Tilman},
	Da = {2010/04/01},
	Doi = {10.1038/nature09009},
	Id = {Baumann2010},
	Isbn = {1476-4687},
	Journal = {Nature},
	Number = {7293},
	Pages = {1301--1306},
	Title = {Dicke quantum phase transition with a superfluid gas in an optical cavity},
	Ty = {JOUR},
	Url = {https://doi.org/10.1038/nature09009},
	Volume = {464},
	Year = {2010}}

@article{helson2023density,
	Author = {Helson, Victor and Zwettler, Timo and Mivehvar, Farokh and Colella, Elvia and Roux, Kevin and Konishi, Hideki and Ritsch, Helmut and Brantut, Jean-Philippe},
	Da = {2023/06/01},
	Doi = {10.1038/s41586-023-06018-3},
	Id = {Helson2023},
	Isbn = {1476-4687},
	Journal = {Nature},
	Number = {7966},
	Pages = {716--720},
	Title = {Density-wave ordering in a unitary Fermi gas with photon-mediated interactions},
	Ty = {JOUR},
	Url = {https://doi.org/10.1038/s41586-023-06018-3},
	Volume = {618},
	Year = {2023}}

@article{zwettler2025cavity,
	Author = {Zwettler, Timo and Marijanovic, Filip and B{\"u}hler, Tabea and Chattopadhyay, Sambuddha and Del Pace, Giulia and Skolc, Luka and Helson, Victor and Uchino, Shun and Demler, Eugene and Brantut, Jean-Philippe},
	Da = {2025/12/08},
	Doi = {10.1038/s41467-025-67184-8},
	Id = {Zwettler2025},
	Isbn = {2041-1723},
	Journal = {Nature Communications},
	Number = {1},
	Pages = {496},
	Title = {Cavity-mediated charge and pair-density waves in a unitary Fermi gas},
	Ty = {JOUR},
	Url = {https://doi.org/10.1038/s41467-025-67184-8},
	Volume = {17},
	Year = {2025}}

@article{zhang2021observation,
author = {Xiaotian Zhang  and Yu Chen  and Zemao Wu  and Juan Wang  and Jijie Fan  and Shujin Deng  and Haibin Wu },
title = {Observation of a superradiant quantum phase transition in an intracavity degenerate Fermi gas},
journal = {Science},
volume = {373},
number = {6561},
pages = {1359-1362},
year = {2021},
doi = {10.1126/science.abd4385},
URL = {https://www.science.org/doi/abs/10.1126/science.abd4385}}

@article{zwettler2025nonequilibrium,
  title = {Nonequilibrium Dynamics of Long-Range Interacting Fermions},
  author = {Zwettler, T. and Del Pace, G. and Marijanovic, F. and Chattopadhyay, S. and B\"uhler, T. and Halati, C.-M. and Skolc, L. and Tolle, L. and Helson, V. and Bolognini, G. and Fabre, A. and Uchino, S. and Giamarchi, T. and Demler, E. and Brantut, J. P.},
  journal = {Phys. Rev. X},
  volume = {15},
  issue = {2},
  pages = {021089},
  numpages = {13},
  year = {2025},
  month = {Jun},
  publisher = {American Physical Society},
  doi = {10.1103/PhysRevX.15.021089},
  url = {https://link.aps.org/doi/10.1103/PhysRevX.15.021089}
}

@article{halati2020numerically,
  title = {Numerically Exact Treatment of Many-Body Self-Organization in a Cavity},
  author = {Halati, Catalin-Mihai and Sheikhan, Ameneh and Ritsch, Helmut and Kollath, Corinna},
  journal = {Phys. Rev. Lett.},
  volume = {125},
  issue = {9},
  pages = {093604},
  numpages = {6},
  year = {2020},
  month = {Aug},
  publisher = {American Physical Society},
  doi = {10.1103/PhysRevLett.125.093604},
  url = {https://link.aps.org/doi/10.1103/PhysRevLett.125.093604}
}

@article{halati2020theoretical,
  title = {Theoretical methods to treat a single dissipative bosonic mode coupled globally to an interacting many-body system},
  author = {Halati, Catalin-Mihai and Sheikhan, Ameneh and Kollath, Corinna},
  journal = {Phys. Rev. Res.},
  volume = {2},
  issue = {4},
  pages = {043255},
  numpages = {20},
  year = {2020},
  month = {Nov},
  publisher = {American Physical Society},
  doi = {10.1103/PhysRevResearch.2.043255},
  url = {https://link.aps.org/doi/10.1103/PhysRevResearch.2.043255}
}

@article{jager2022lindblad,
  title = {Lindblad Master Equations for Quantum Systems Coupled to Dissipative Bosonic Modes},
  author = {J\"ager, Simon B. and Schmit, Tom and Morigi, Giovanna and Holland, Murray J. and Betzholz, Ralf},
  journal = {Phys. Rev. Lett.},
  volume = {129},
  issue = {6},
  pages = {063601},
  numpages = {7},
  year = {2022},
  month = {Aug},
  publisher = {American Physical Society},
  doi = {10.1103/PhysRevLett.129.063601},
  url = {https://link.aps.org/doi/10.1103/PhysRevLett.129.063601}
}

@article{halati2025controlling,
  title = {Controlling the Dynamics of Atomic Correlations via the Coupling to a Dissipative Cavity},
  author = {Halati, Catalin-Mihai and Sheikhan, Ameneh and Morigi, Giovanna and Kollath, Corinna},
  journal = {Phys. Rev. Lett.},
  volume = {134},
  issue = {7},
  pages = {073604},
  numpages = {7},
  year = {2025},
  month = {Feb},
  publisher = {American Physical Society},
  doi = {10.1103/PhysRevLett.134.073604},
  url = {https://link.aps.org/doi/10.1103/PhysRevLett.134.073604}
}

@article{marijanovic_quench_2026,
	title = {Quench {Instabilities} of a {Strongly} {Interacting} {Quantum} {Gas} in an {Optical} {Cavity}},
	volume = {136},
	issn = {0031-9007, 1079-7114},
	url = {https://link.aps.org/doi/10.1103/w2vx-t1mr},
	doi = {10.1103/w2vx-t1mr},
	language = {en},
	number = {19},
	urldate = {2026-06-29},
	journal = {Physical Review Letters},
	author = {Marijanović, Filip and Chattopadhyay, Sambuddha and Skolc, Luka and Zwettler, Timo and Halati, Catalin-Mihai and Jäger, Simon B. and Giamarchi, Thierry and Brantut, Jean-Philippe and Demler, Eugene},
	month = may,
	year = {2026},
	pages = {193401},
}

@article{schuler_nessi_2020,
	title = {{NESSi}: {The} {Non}-{Equilibrium} {Systems} {Simulation} package},
	volume = {257},
	issn = {00104655},
	shorttitle = {{NESSi}},
	url = {http://arxiv.org/abs/1911.01211},
	doi = {10.1016/j.cpc.2020.107484},
	urldate = {2026-07-10},
	journal = {Computer Physics Communications},
	author = {Schüler, Michael and Golež, Denis and Murakami, Yuta and Bittner, Nikolaj and Hermann, Andreas and Strand, Hugo U. R. and Werner, Philipp and Eckstein, Martin},
	month = dec,
	year = {2020},
	pages = {107484},
}

@article{picano_quantum_2021,
	title = {A quantum {Boltzmann} equation for strongly correlated electrons},
	volume = {104},
	issn = {2469-9950, 2469-9969},
	url = {http://arxiv.org/abs/2101.09037},
	doi = {10.1103/PhysRevB.104.085108},
	number = {8},
	urldate = {2026-01-15},
	journal = {Physical Review B},
	author = {Picano, Antonio and Li, Jiajun and Eckstein, Martin},
	month = aug,
	year = {2021},
	pages = {085108},
}

@article{eckstein_thermalization_2009,
	title = {Thermalization after an {Interaction} {Quench} in the {Hubbard} {Model}},
	volume = {103},
	url = {https://link.aps.org/doi/10.1103/PhysRevLett.103.056403},
	doi = {10.1103/PhysRevLett.103.056403},
	number = {5},
	urldate = {2026-09-11},
	journal = {Physical Review Letters},
	publisher = {American Physical Society},
	author = {Eckstein, Martin and Kollar, Marcus and Werner, Philipp},
	month = jul,
	year = {2009},
	pages = {056403},
}

@article{RevModPhys.86.779,
  title = {Nonequilibrium dynamical mean-field theory and its applications},
  author = {Aoki, Hideo and Tsuji, Naoto and Eckstein, Martin and Kollar, Marcus and Oka, Takashi and Werner, Philipp},
  journal = {Rev. Mod. Phys.},
  volume = {86},
  issue = {2},
  pages = {779--837},
  numpages = {59},
  year = {2014},
  month = {Jun},
  publisher = {American Physical Society},
  doi = {10.1103/RevModPhys.86.779},
  url = {https://link.aps.org/doi/10.1103/RevModPhys.86.779}
}

@article{picano2025quantum,
  title = {Quantum Thermalization via Travelling Waves},
  author = {Picano, Antonio and Biroli, Giulio and Schir\`o, Marco},
  journal = {Phys. Rev. Lett.},
  volume = {134},
  issue = {11},
  pages = {116503},
  numpages = {7},
  year = {2025},
  month = {Mar},
  publisher = {American Physical Society},
  doi = {10.1103/PhysRevLett.134.116503},
  url = {https://link.aps.org/doi/10.1103/PhysRevLett.134.116503}
}

@misc{picano2025heatingdynamicscorrelatedfermions,
	title = {Heating {Dynamics} of {Correlated} {Fermions} under {Dephasing}},
	copyright = {Creative Commons Attribution 4.0 International},
	url = {https://arxiv.org/abs/2507.21804},
	doi = {10.48550/arxiv.2507.21804},
	language = {en},
	urldate = {2026-09-14},
	publisher = {arXiv},
	author = {Picano, Antonio and Vanhoecke, Matthieu and Schirò, Marco},
	year = {2025},
}
\end{appendix}

\end{document}